\documentclass[preprints,article,accept,moreauthors,pdftex]{Definitions/mdpi} 

\firstpage{1} 
\pubvolume{xx}
\issuenum{1}
\articlenumber{5}
\pubyear{2019}
\copyrightyear{2019}
\history{Received: date; Accepted: date; Published: date}

\usepackage{graphicx}
\usepackage{subfigure}
\usepackage{amssymb}

\Title{Effect of Single-ion Anisotropy on Magnetocaloric Properties of Frustrated Spin-$s$ Ising Nanoclusters}

\Author{Mariia Mohylna $^{1}$ and Milan \v{Z}ukovi\v{c} $^{1}$\orcidA{}}

\AuthorNames{Mariia Mohylna and Milan \v{Z}ukovic\v{c}}

\address{%
$^{1}$ \quad Institute of Physics, Faculty of Science, P. J. \v{S}af\'arik University, Park Angelinum 9, 041 54 Ko\v{s}ice, Slovakia; milan.zukovic@upjs.sk}

\corres{Correspondence: milan.zukovic@upjs.sk}

\abstract{Effects of a single-ion anisotropy on magnetocaloric properties of selected spin-$s\geq 1$ antiferromagnetic Ising clusters with frustration-inducing triangular geometry are studied by exact enumeration. It is found that inclusion of the single-ion anisotropy parameter $D$ can result in a much more complex ground-state behavior, which is also reflected in a magnetocaloric effect (MCE) at finite temperatures. For negative $D$ (easy-plane anisotropy) with increasing $s$ the ground-state magnetization as a function of the external field gradually shows increasing number of plateaus of various heights. Except for the cases of integer $s$ with $D < D_0 \leq 0$, the first magnetization plateau is of non-zero height. This property facilitates an enhanced MCE in the adiabatic demagnetization process in the form of an abrupt decrease in temperature as the magnetic field vanishes to zero. The cooling rate can be considerably enhanced in the systems with larger $s$ and $D>0$ (easy-axis anisotropy), albeit its dependence on these parameters is strongly dependent on the cluster geometry. From the studied systems more favorable conditions for observing a giant MCE were found in the 2CS cluster, consisting of two corner-sharing tetrahedra, the experimental realization of which could be technologically used for efficient refrigeration to ultra-low temperatures.}

\keyword{Ising antiferromagnet; geometrical frustration; nanocluster; higher spin; single-ion anisotropy; exact enumeration; magnetocaloric effect}

\begin{document}


\section{Introduction}

Molecular nanomagnets often occur in nature and nowadays owing to new technologies can also be artificially prepared in a highly controlled manner~\cite{shen02,loun07,pede14}. They can be viewed as zero-dimensional magnetic structures composed of small spin clusters, which are magnetically isolated from the environment. In spite of their simplicity, they have drawn considerable interest~\cite{kahn93,gatt06,furr13} mainly due to their potential applications in data storage, quantum computing, molecule-based spintronics devices as well as magnetic cooling technology. The last one is related to their ability to display a very large magnetocaloric effect (MCE) at low temperatures~\cite{spic01,evan06,mano07,mano08}. 

Large MCE can result from various features displayed by these magnetic structures. For example, in high-spin molecular nanomagnets macroscopic degeneracy of the molecular spin states at low temperatures translates to excess of magnetic entropy and consequently in an increased MCE~\cite{evan05,shaw07}. The spin degeneracy gradually increases with the spin number and further increase can be achieved by designing very weak magnetic links between the single-ion spin centers~\cite{evan09}. Another aspect that can result in a high degeneracy at low temperatures is geometrical frustration. A number of geometrically frustrated antiferromagnetic systems are known to display no phase transition and finite entropy even in ground state. Compared to paramagnetic salts, which are considered standard refrigerant materials for magnetic cooling, they were theoretically predicted~\cite{zhit03,schn07} and experimentally confirmed, for example on a pyrochlore lattice compound $\rm{Gd}_2\rm{Ti}_2\rm{O}_7$~\cite{sosi05}, to show more than an order of magnitude bigger rate of the temperature decrease due to the varying of the external magnetic field.

In this respect of particular interest are low-dimensional geometrically frustrated systems, such as zero-dimensional molecular nanomagnets. It is worthwhile mentioning that the first direct evidence of sub-Kelvin cooling achieved with a molecular nanomagnet was recently observed in geometrically frustrated ${\rm Gd}_7$~\cite{shar14,pine16}. The experiment revealed isentropes with a rich structure, which were very well reproduced theoretically by considering a simple Heisenberg spin Hamiltonian with only exchange ~\cite{shar14} or including also intra-molecular dipolar interactions~\cite{pine16}. The observed isentropes were concluded to be a direct manifestation of the trigonal antiferromagnetic net structure, which triggered further experimental~\cite{oren18,fitt19} and theoretical~\cite{stre15,karl16,milla1,zuko14a,zuko15,zuko18a,mohy19,hald20,kowa20,szal20a,szal20b} studies of mainly frustration-enhanced MCE in finite systems of various shapes and topologies. They included geometrically frustrated Ising spin clusters with shapes of regular polyhedra~\cite{stre15,karl16} or triangular geometry~\cite{milla1,zuko14a,zuko15,zuko18a,mohy19}. It turns out that magnetocaloric properties are highly sensitive to the cluster geometry. In particular, for low-temperature cooling spin clusters which lack zero-magnetization plateaus in an applied magnetic field at zero temperature, such as the octahedron and dodecahedron from the regular polyhedra~\cite{stre15} and most of the studied clusters composed of triangular plaquettes~\cite{zuko15,zuko18a,mohy19}, were identified as promising candidates since they exhibited a giant magnetocaloric effect in the process of adiabatic demagnetization.

Nevertheless, besides the shape and topology there are other properties, such as spin, anisotropy (both exchange and single-ion), higher-order interactions, surface effects, etc., that can affect the spin system properties~\cite{efre06,efre08,huch11,shap02} including their magnetocaloric behavior~\cite{shar13,yuks18,zad18,zad20}. In the present paper we focus on two types of the geometrically frustrated triangle-based Ising nanoclusters with antiferromagnetic interaction that have shown favorable low-temperature MCE for $s=1/2$~\cite{zuko15,mohy19} and generalize them by considering a higher spin $s \geq 1$ and inclusion of the single-ion anisotropy. In particular, we study a simple triangular (T) cluster~\cite{zuko15} and a cluster composed of two corner-sharing (2CS) tetrahedra~\cite{mohy19}. We demonstrate that their magnetocaloric properties strongly depend on both the spin as well as the sign and magnitude of the single-ion anisotropy and identify some particular arrangements that can lead to a giant MCE in a small or even vanishing magnetic field in the adiabatic demagnetization process.

\section{Model and method}
\subsection{Model}
We consider an Ising spin-$s$ cluster with triangular geometry in an external magnetic field described by the Hamiltonian 

\begin{equation} \label{Hamilt}
\mathcal{H}=-J\sum_{\langle i,j\rangle}s_i s_j - D\sum_i s_i^2 - h\sum_i s_i,
\end{equation}
where $s_i=-s,-s+1,\hdots +s-1,+s$ is the spin variable on the $i$-th site of the cluster, the summation $\langle i,j\rangle$ runs over all nearest-neighbor pairs, $J<0$ is an antiferromagnetic exchange interaction, $D$ is a single-ion anisotropy parameter, and $h$ is the external magnetic field. 

We focus on two types of small spin clusters with simple shapes composed of elementary triangular plaquettes, which have been shown to display favorable magnetocaloric properties for the spin-$1/2$ models. Namely, we consider the T cluster composed of three spins located on a equilateral triangle, i.e., the elementary triangular plaquette, and the 2CS cluster composed of seven spins located on two corner-sharing elementary regular tetrahedra  (see insets in Figures~\ref{fig:triangle_m_1} and~\ref{fig:2CS_m_1}).
  
\subsection{Method of exact enumeration}
\subsubsection{Ground state}
Considering the small numbers of spins in the respective clusters it is feasible to explore the entire state space of the systems even for larger spin values and thus exactly determine the density of states. Consequently one can obtain all the thermodynamic quantities of interest. In order to determine the ground state (GS) in a broad parameter space one needs to identify the spin states $\{s_1,s_2,\ldots,s_N\}$ which minimize the energy functional (\ref{Hamilt}). For different parameters there might be multiple configurations corresponding to the same minimum energy and by counting their number one can obtain the GS degeneracy $W$. The quantities of interest include magnetization, which can be obtained as $M_{k} = \sum_{i=1}^N s_{i}$, for $k=1,2, \ldots W$, for each identified GS configuration. Then, the total magnetization per spin can be calculated as:

\begin{equation} \label{namtot}
m=\frac{1}{WN}\sum_{k=1}^W M_k
\end{equation}
and the entropy per spin as

\begin{equation} \label{entr}
\frac{S}{Nk_B}=\frac{1}{N}\ln W,
\end{equation}
where $k_B$ is the Boltzmann constant.
 
\subsubsection{Thermodynamic and magnetocaloric properties}
The exact enumeration method can be extended to calculations of various thermodynamic quantities at finite temperatures as functions of the model parameters. Below we focus on presenting the calculated quantities for a fixed $D$ in the $T-h$ parameter plane. For that purpose one needs to calculate the density of states $g(M,E)$ as a function of the magnetization $M$ and the energy $E=\sum_{\langle i,j\rangle}s_i s_j-D\sum_{i}s_i^2$. Consequently one can obtain the partition function, as a function of the model parameters $T$ and $h$, as follows

\begin{equation} \label{stsum}
Z(T,h)=\sum_{M,E}g(M,E)e^{-\frac{E-hM}{T}}.
\end{equation}
Having obtained the partition function, we can easily evaluate any thermodynamic quantity of interest. Namely, a mean value of any thermodynamic function $A$ can be calculated as a function of the temperature and the field as

\begin{equation} \label{stred}
\langle A(T,h)\rangle =\frac{\sum_{M,E} A g(M,E)e^{-\frac{E-hM}{T}}}{Z(T,h)}.
\end{equation}
In particular, the magnetization per spin is calculated as 

\begin{equation} \label{magnsr}
m(T,h)=\frac{\langle M(T,h)\rangle}{N}
\end{equation} 
and the magnetic entropy density (entropy per spin) in the form

\begin{equation} \label{entrmain}
\frac{S(T,h)}{Nk_B}=\frac{U(T,h)-F(T,h)}{NT},
\end{equation}
where $U(T,h)=\langle E(T,h)-hM(T,h) \rangle$ is the enthalpy and $F(T,h)=-T \ln Z(T,h)$ is the free energy.

Particularly the behavior of the above quantities, i.e., the magnetization and the entropy as functions of the temperature and the field, characterizes magnetocaloric properties of the studied systems. One important quantity is the isothermal entropy density change, which can be obtained as

\begin{equation} \label{entr_change}
\Delta S(T,\Delta h)/Nk_B =(S_f-S_i)/Nk_B. 
\end{equation}
It corresponds to the change of the initial entropy $S_i$ to the final value $S_f$ when the field changes from the initial $h_i$ to the final value $h_f$, i.e., by $\Delta h = h_f-h_i$, at the constant temperature. Another quantity that characterizes magnetocaloric properties is the adiabatic magnetic cooling rate $C_r$, which is defined as

\begin{equation} \label{cool_rate}
C_r =-\frac{(\partial S/\partial h)_T}{(\partial S/\partial T)_h}, 
\end{equation}
where the lower indices $T$ and $h$ signify the constant temperature and field, respectively. Below, the temperature, the magnetic field and the single-ion anisotropy will be presented in dimensionless units $k_{B}T/|J|$, $h/|J|$, and $D/|J|$, respectively.

\section{Results}

\subsection{Ground state}
\subsubsection{T cluster}
\begin{figure}[t!]
\centering
\subfigure{\includegraphics[width=8 cm]{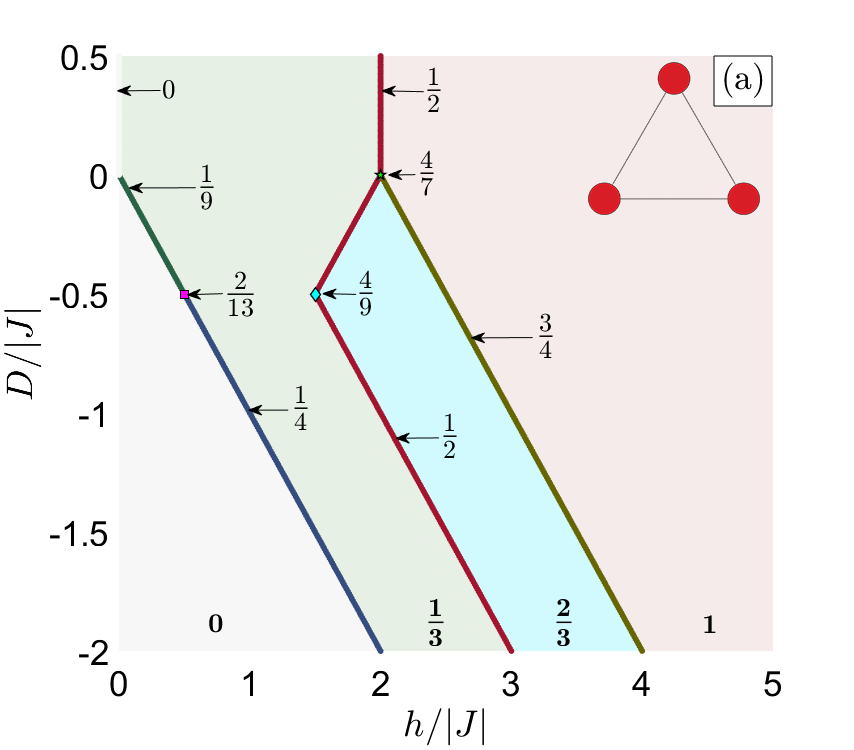}\label{fig:triangle_m_1}}\hspace{-5mm}
\subfigure{\includegraphics[width=8 cm]{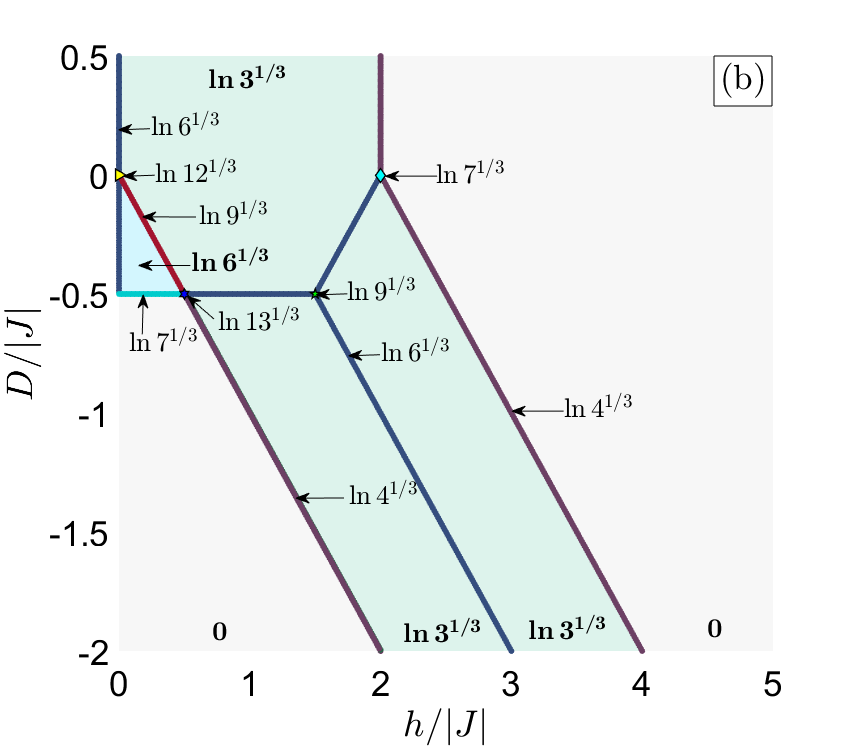}\label{fig:triangle_entr_1}}\\
\subfigure{\includegraphics[width=8 cm]{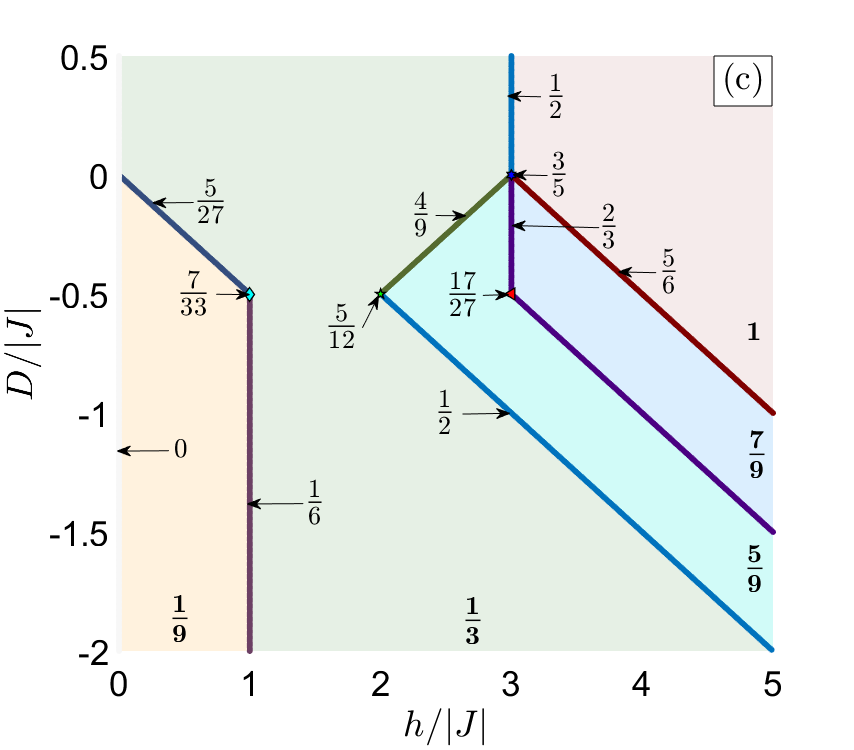}\label{fig:triangle_m_1.5}}\hspace{-5mm}
\subfigure{\includegraphics[width=8 cm]{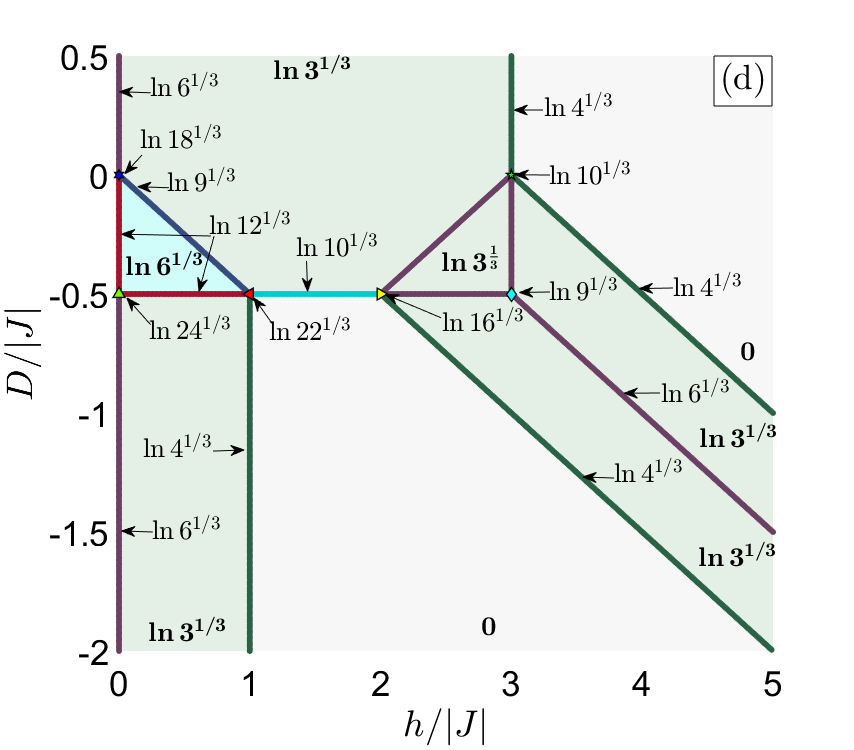}\label{fig:triangle_entr_1.5}}
\caption{Ground state values of (a,c) the normalized magnetization $m/m_{sat}$ and (b,d) the entropy density $S/Nk_B$ of the T cluster with (a,b) $s=1$ and (c,d) $s=3/2$, in the $(h/|J|,D/|J|)$ parameter plane. Boldface (plane) values correspond to the interior (boundaries) of the respective phases. The inset in (a) shows the T cluster.}\label{fig:gs_T}
\end{figure}

In Figure~\ref{fig:gs_T} we present the ground-state magnetization (left column) and entropy (right column) of the T cluster in the $(h/|J|,D/|J|)$ parameter plane, with the spins $s=1$ (first row), representing the integer spin values, and $s=3/2$ (second row), representing the half-integer spin values. 

For $D/|J|>0$ we obtain similar behavior of both quantities for all the spin values (including the cases with $s>3/2$ not shown here), which reproduces the behavior of the spin-$1/2$ model. Namely, an arbitrarily small magnetic field increases the magnetization from zero to one third of its saturation value, $m_{sat}=s$, within $0<h/|J|<2s$ and decreases the corresponding entropy density from $S/Nk_B=\ln(6)/3$ to $\ln(3)/3$. Right at the saturation field $h_{sat}/|J|=2s$ the entropy density shows a spike reaching the value $\ln(4)/3$ and the corresponding normalized magnetization equals to $1/2$. We also note that both the magnetization and entropy values at the critical fields, at which both quantities show discontinuities, generally differ from the values inside the plateaus due to the coexistence of states with different degeneracies. For $h/|J|>2s$ the system becomes fully polarized, the magnetization fully saturated and the entropy density drops to zero. For $D/|J|=0$ the behavior is similar to the $D/|J|>0$ case, except for the gradual increase of the entropy density at $h/|J|=0$ and $h/|J|=h_{sat}/|J|$ with the increasing spin value. In particular, in zero field the degeneracy can be expressed as $W=2s(2^N-2)$ and the at the saturation field as $W=2s(2^{N-1}-1)+1$.

\begin{figure}[t!]
\centering
\subfigure{\includegraphics[width=8 cm]{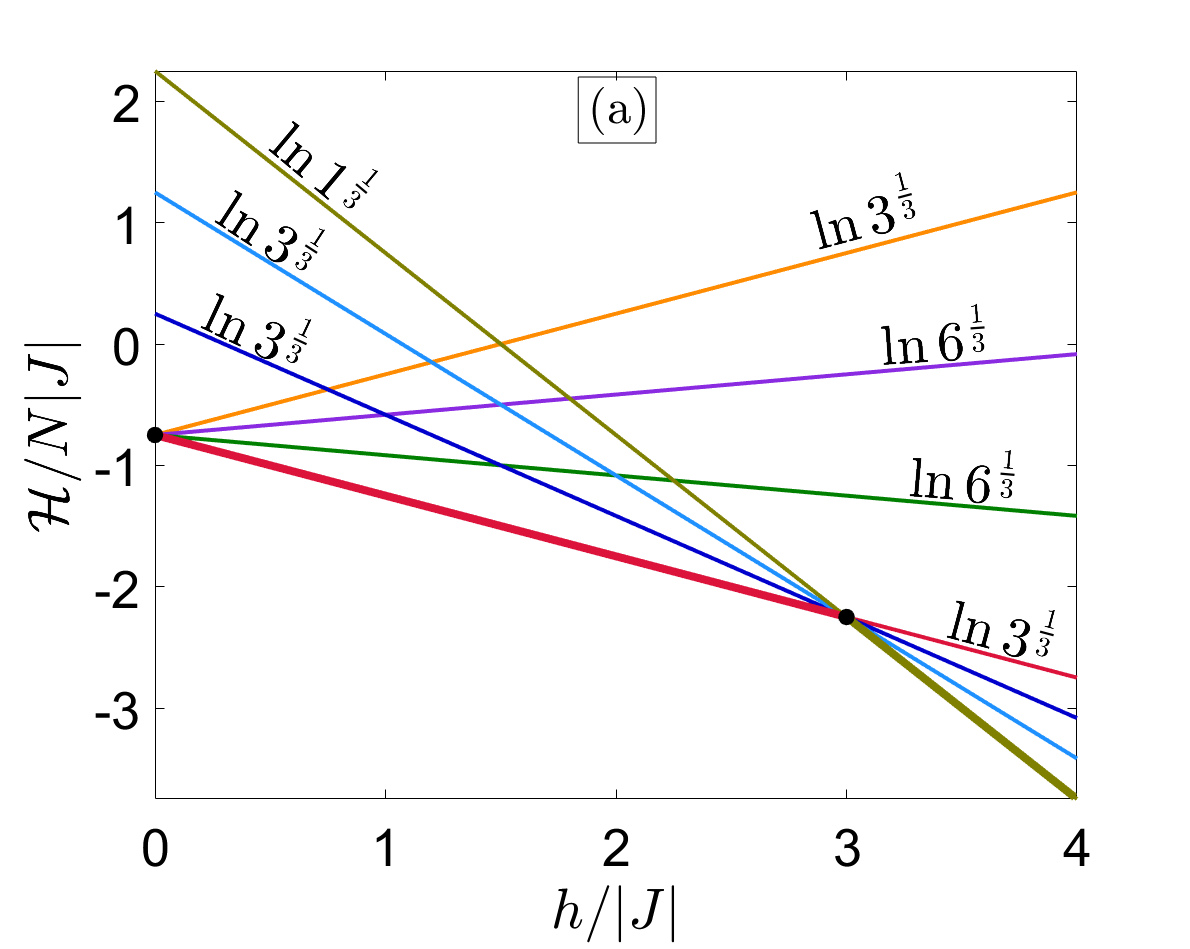}\label{fig:T_lvl_S1-5_A0}}\hspace{-5mm}
\subfigure{\includegraphics[width=8 cm]{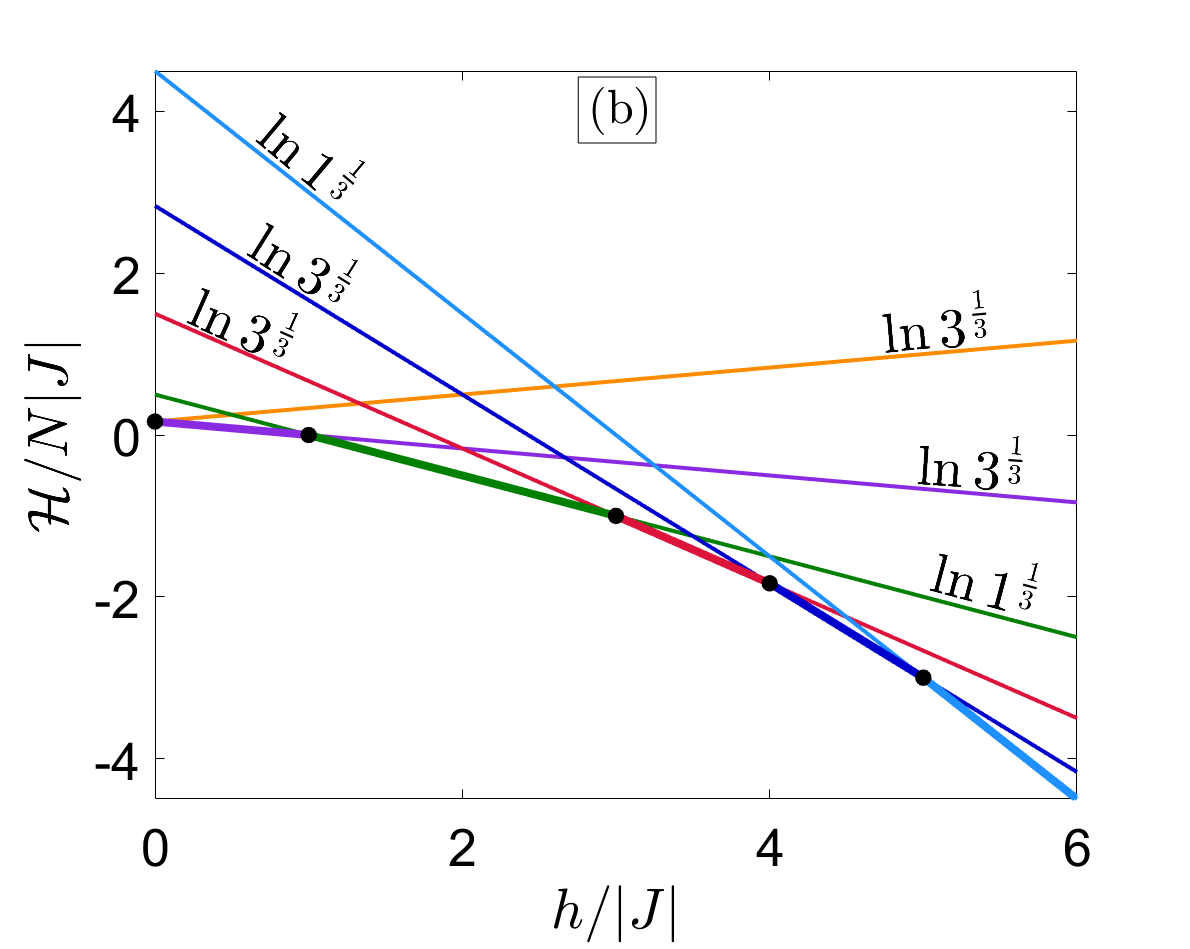}\label{fig:T_lvl_S1-5_A-1}}
\caption{Energy levels of the T cluster with $s=3/2$ for (a) $D/|J|=0$ and (b) $D/|J|=-1$ as functions of the field. The thick lines correspond to the ground-state energy levels within given field ranges. The listed entropy density values show degeneracies of the respective energy levels.}\label{fig:en_lev}
\end{figure}

Negative values of $D/|J|$ lead to a more complex behavior. The one-third magnetization plateau splits for $s=1$ into three ($0,1/3,2/3$) and for $s=3/2$ into four ($1/9,1/3,5/9,7/9$) intermediate plateaus separated by the critical fields $h_{ci}/|J|$, where $i=1,\hdots,i_{max}$ and $i_{max}$ is the number of the intermediate plateaus, before the fully saturated state is reached\footnote{In the absence of a zero-magnetization plateau the first critical field corresponds to $h_{c1}/|J|=0$.}. Down to $D/|J|=-0.5$ the decreasing $D/|J|$ makes the one-third magnetization plateau shrink at the cost of the newly created plateaus. Further decrease of $D/|J|$ makes the widths of some plateaus increase while some others remain unchanged. Note that the plateaus' heights increase with the spin $s$ by the increments $(3s)^{-1}$. In the case of integer (half-integer) $s$ they start increasing from zero at the first critical field corresponding to $h_{c1}/|J| = -D/|J|$ ($h_{c1}/|J| = 0$). Thus, from the MCE point of view more interesting are the cases of the half-integer $s$, showing no zero-magnetization plateau. Moreover, the half-integer spin models display for any $D/|J|<-0.5$ a relatively large entropy change when the field is reduced to zero from a comparatively small initial value of $ h_i \gtrsim 1$. For better visualization of the energy-level crossings and degeneracy changes at the critical fields, in Figure~\ref{fig:en_lev} we present energy levels of the T cluster with $s=3/2$, as functions of the field, for the isotropic case (\ref{fig:T_lvl_S1-5_A0}) and $D/|J|=-1$ (\ref{fig:T_lvl_S1-5_A-1}). The listed entropy density values show degeneracies of the respective energy levels and the ground-state energy levels within the respective field ranges are highlighted by the thick lines.

\subsubsection{2CS cluster}

\begin{figure}[t!]
\centering
\subfigure{\includegraphics[width=8 cm]{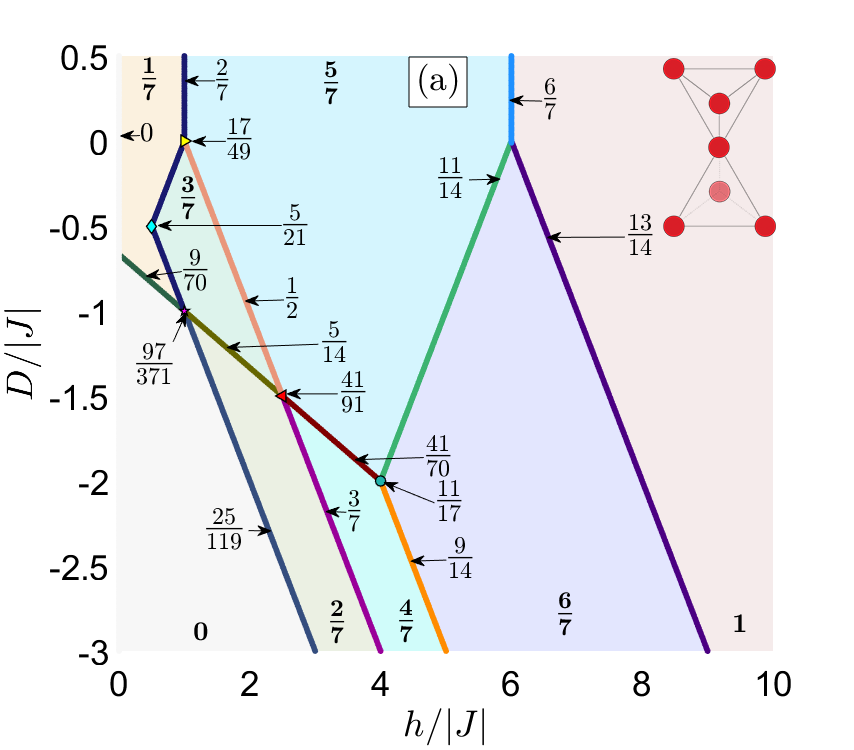}\label{fig:2CS_m_1}}\hspace{-5mm}
\subfigure{\includegraphics[width=8 cm]{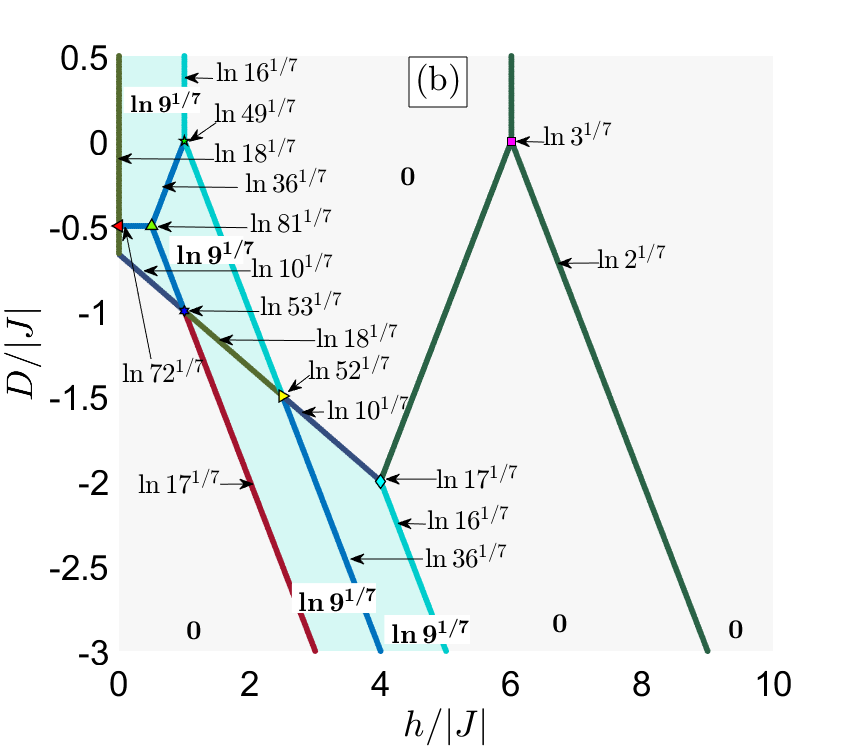}\label{fig:2CS_entr_1}}\\
\subfigure{\includegraphics[width=8 cm]{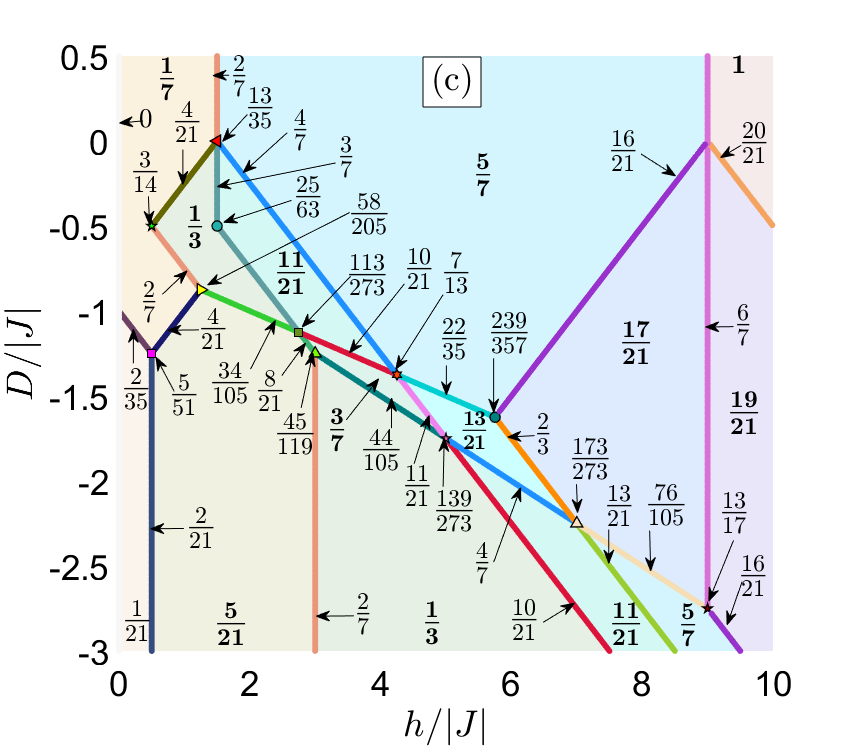}\label{fig:2CS_m_1.5}}\hspace{-5mm}
\subfigure{\includegraphics[width=8 cm]{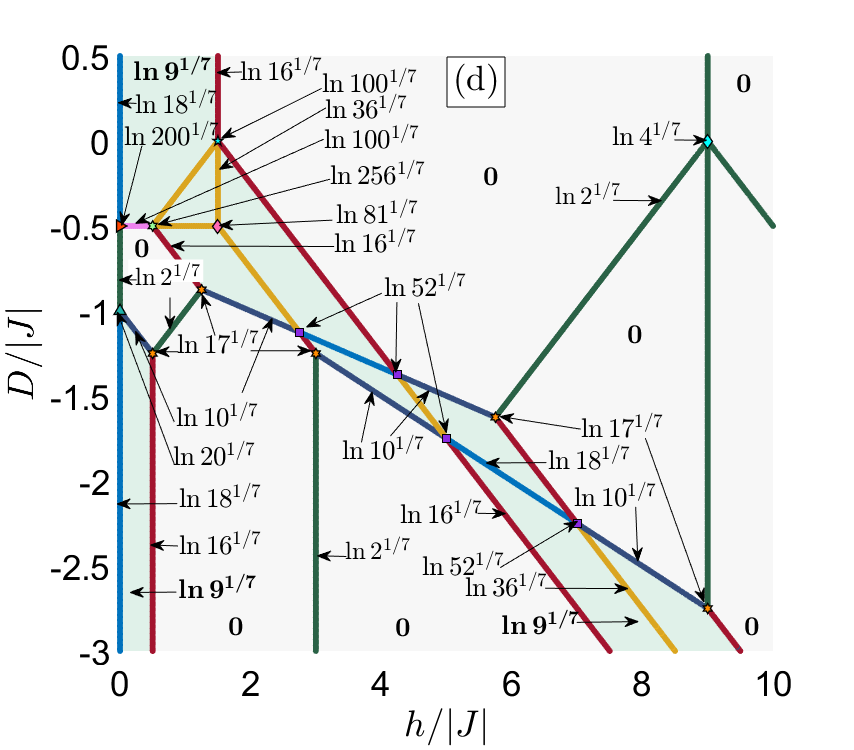}\label{fig:2CS_entr_1.5}}
\caption{Ground state values of (a,c) the normalized magnetization $m/m_{sat}$ and (b,d) the entropy density $S/Nk_B$ of the 2CS cluster with (a,b) $s=1$ and (c,d) $s=3/2$, in the $(h/|J|,D/|J|)$ parameter plane. Boldface (plane) values correspond to the interior (boundaries) of the respective phases. The inset in (a) shows the 2CS cluster.}\label{fig:gs_2CS}
\end{figure}

Also in the case of the 2SC cluster (Figure~\ref{fig:gs_2CS}), for $D/|J|>0$ the behavior is the same as in the spin-$1/2$ model. Specifically, the magnetization shows two plateaus: the first one of the height $m/m_{sat}=1/7$, which spans within $0<h/|J|<s$, is followed by the second one of the height $5/7$, extending within $s<h/|J|<6s$. The entropy density decreases from $S/Nk_B=\ln(18)/7$ at $h/|J|=0$ to $\ln(9)/7$ within $0<h/|J|<s$. Right at $h/|J|=s$ it increases again to $\ln(16)/7$ and completely vanishes for $h/|J|>s$, except at $h/|J|=6s$ at which it acquires the value of $\ln(2)/7$. For $D/|J|=0$ the entropy density at $h/|J|=0$ remains the same as for $D/|J|>0$, i.e., $\ln(18)/7$, but it increases at $h/|J|=s$ ($h/|J|=6s$) to the values $\ln(49)/7$ ($\ln(3)/7$) for $s=1$ and $\ln(100)/7$ ($\ln(4)/7$) for $s=3/2$.

For $D/|J|<0$, similarly to the T cluster, the intermediate magnetization plateaus split into several shorter steps of irregular heights, the number of which increases with the increasing spin. The corresponding magnetization and entropy density values inside the respective plateaus (bold face) as well as on the boundaries (plain) are displayed in Figure~\ref{fig:gs_2CS}. From the MCE perspective again the most interesting are the cases of the half-integer spin, which lack the zero-magnetization plateau for arbitrary $D/|J|$. Furthermore, for any $D/|J|<-1$ they show a relatively large entropy change when the field is reduced from a small initial value of $ h_i \gtrsim 0.5$ or even less close to $D/|J|=-1$.

\subsection{Finite temperatures}
\subsubsection{T cluster}

\begin{figure}[t!]
\centering
\subfigure{\includegraphics[width=5 cm]{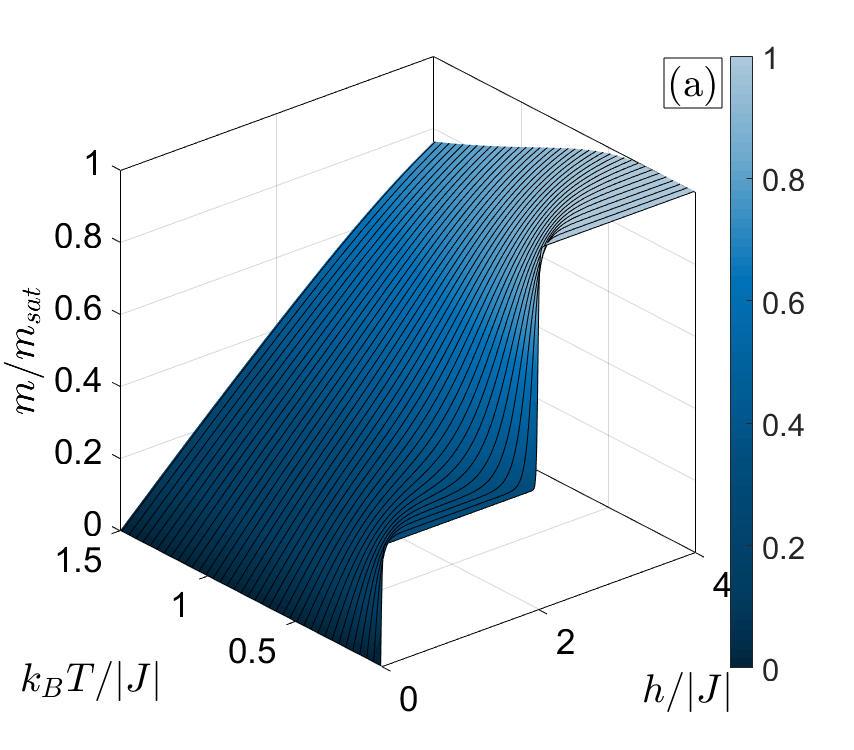}\label{fig:trianglemagn_S1_A0}}
\subfigure{\includegraphics[width=5 cm]{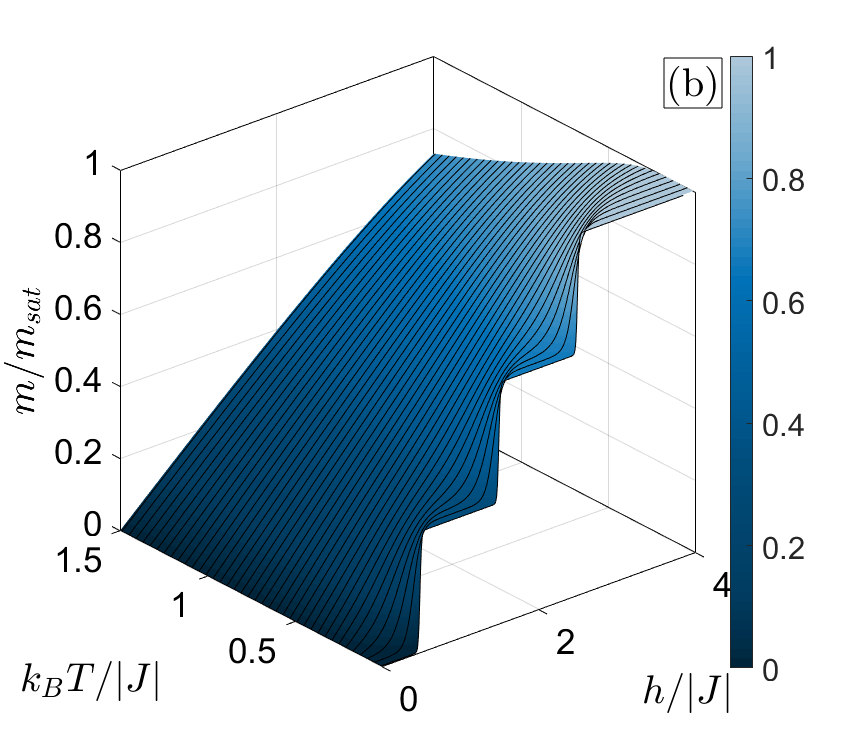}\label{fig:trianglemagn_S1_A-0.5}}
\subfigure{\includegraphics[width=5 cm]{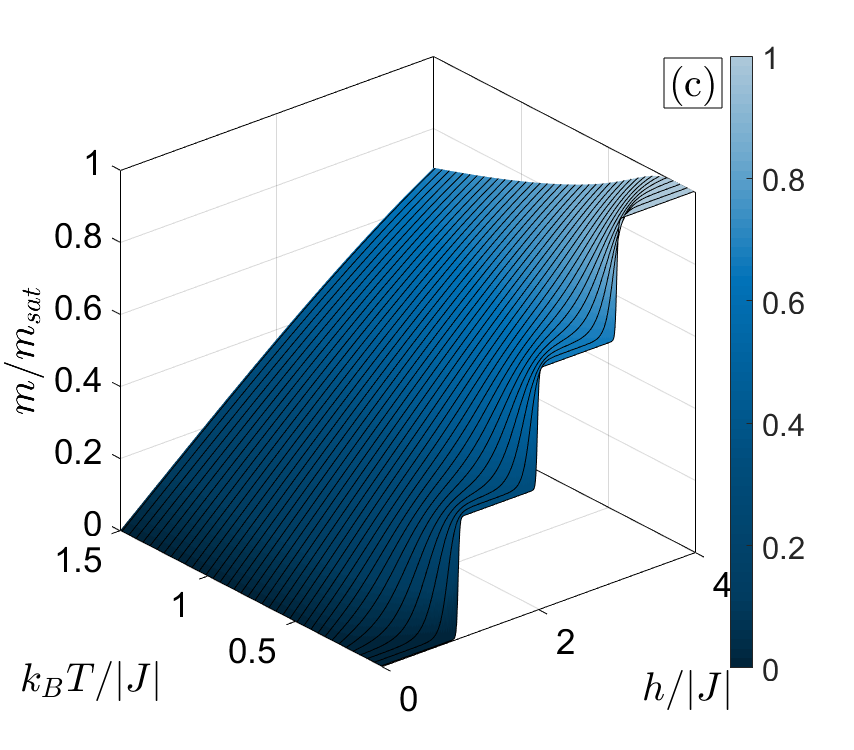}\label{fig:trianglemagn_S1_A-1}}\\
\subfigure{\includegraphics[width=5 cm]{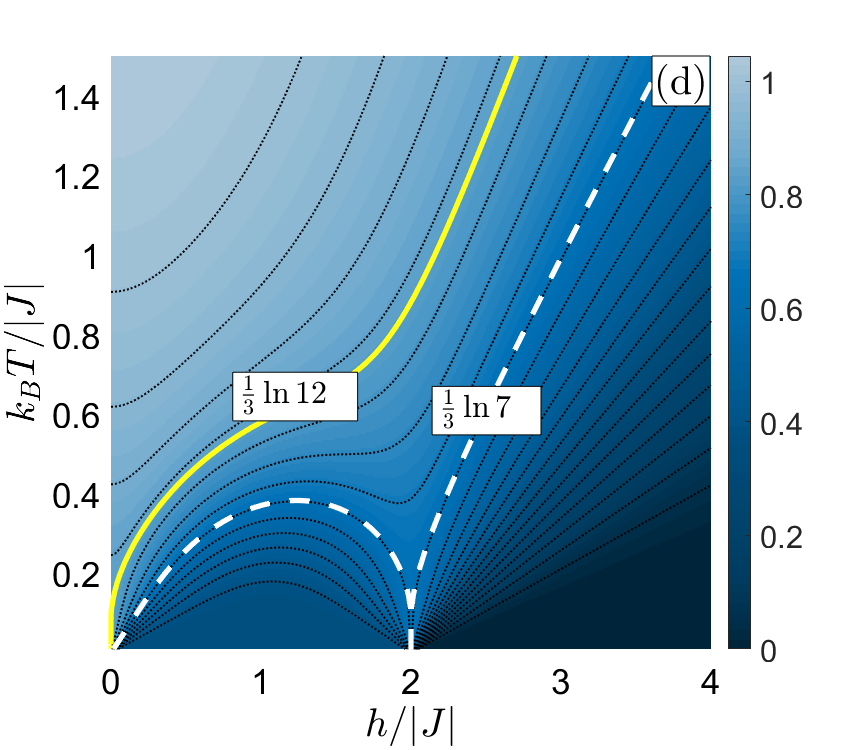}\label{fig:triangleentr_top_S1_A0}}
\subfigure{\includegraphics[width=5 cm]{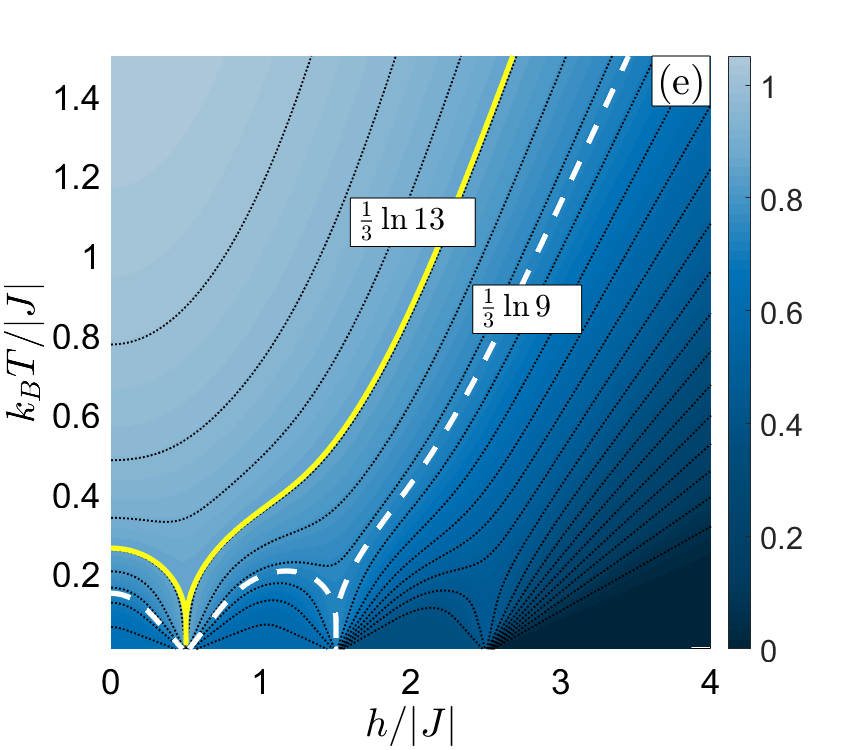}\label{fig:triangleentr_top_S1_A-0.5}}
\subfigure{\includegraphics[width=5 cm]{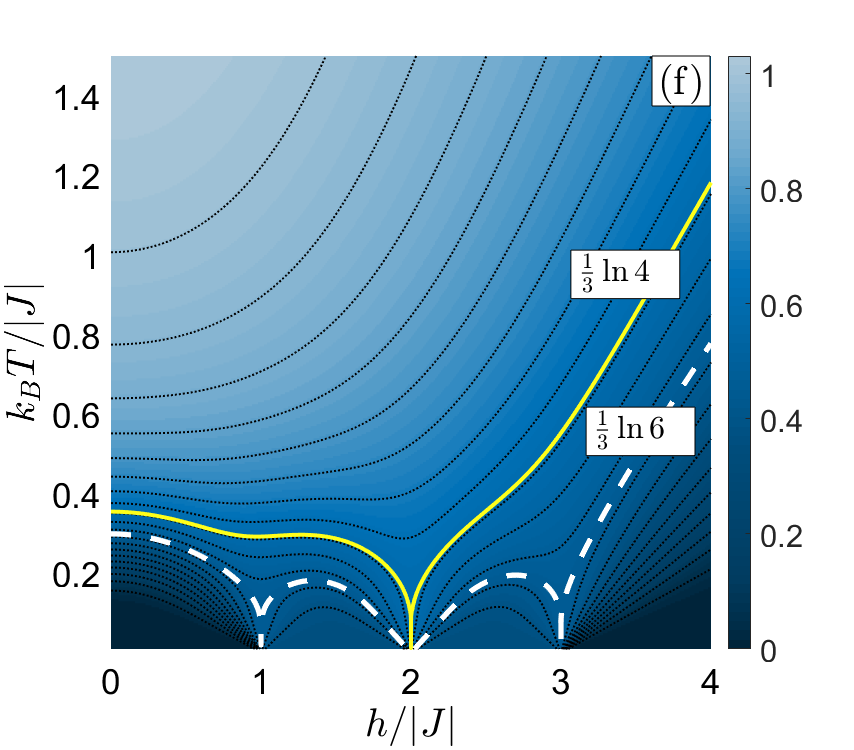}\label{fig:triangleentr_top_S1_A-1}}\\
\subfigure{\includegraphics[width=5 cm]{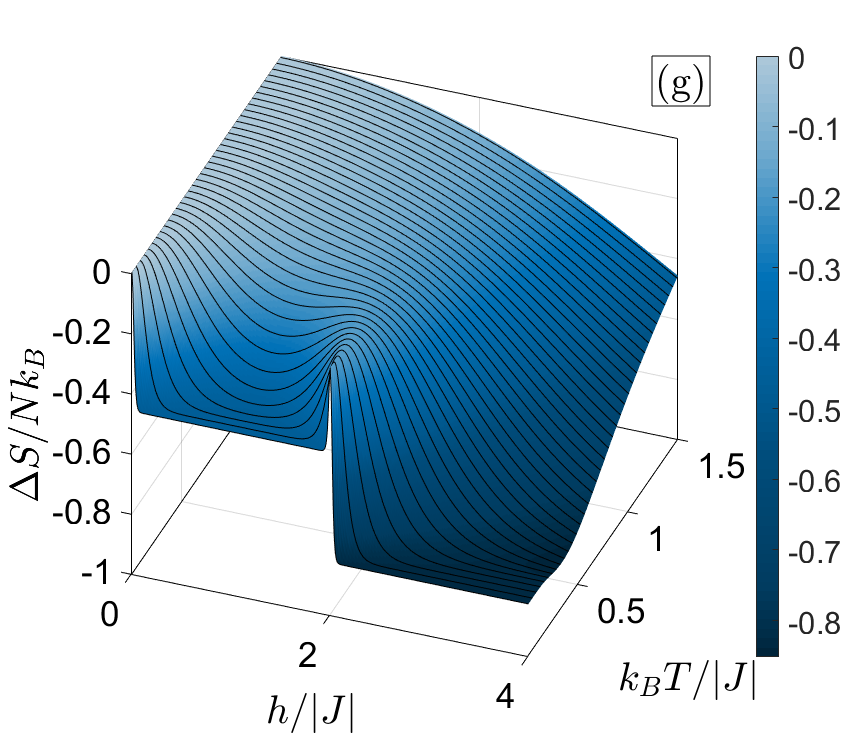}\label{fig:triangleentr_delta_S1_A0}}
\subfigure{\includegraphics[width=5 cm]{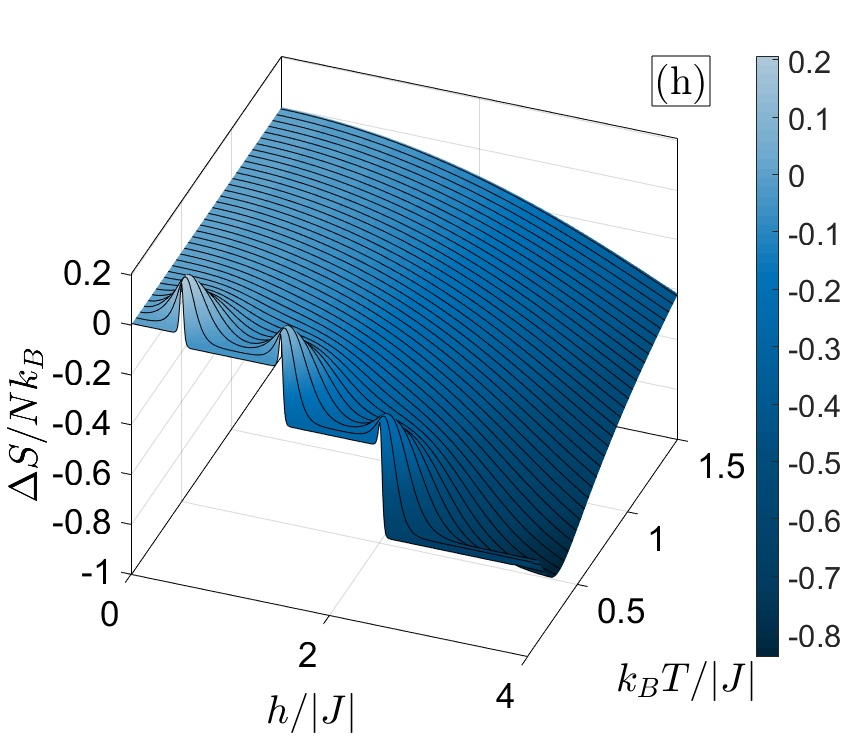}\label{fig:triangleentr_delta_S1_A-0.5}}
\subfigure{\includegraphics[width=5 cm]{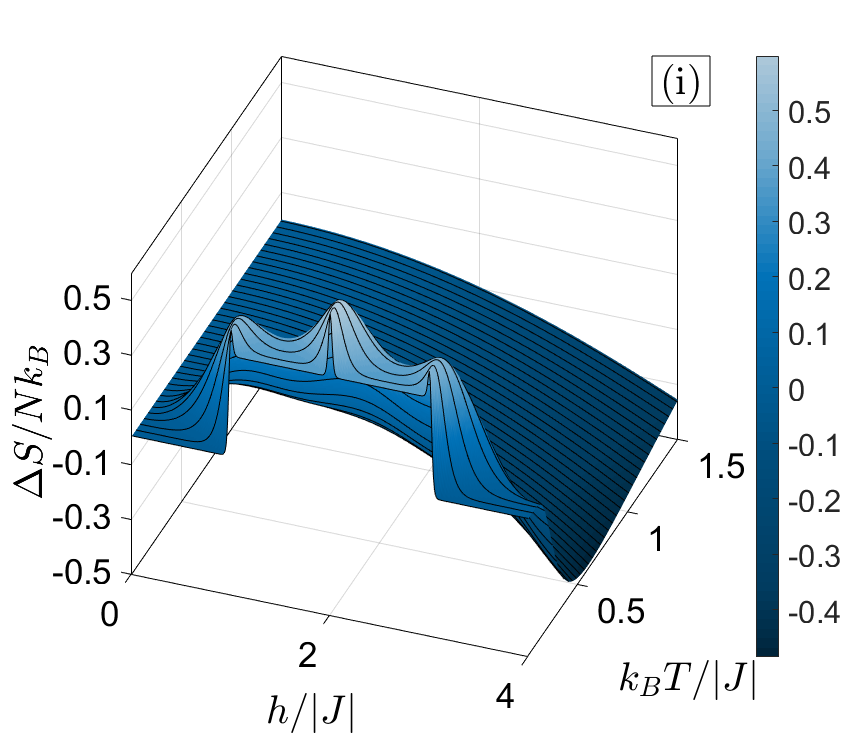}\label{fig:triangleentr_delta_S1_A-1}}
\caption{Magnetization (top row), entropy density (middle row), and isothermal entropy density change (bottom row) in $k_{B}T/|J|-h/|J|$ plane, for T cluster with $s=1$ and $D/|J|=0$ (first column), $D/|J|=-0.5$ (second column), and $D/|J|=-1$ (last column). The highlighted isentropes in the middle row correspond to the GS residual entropies at the first two critical fields $h_{c1}/|J|$ (solid yellow) and $h_{c2}/|J|$ (dashed white).}\label{fig:ft_T_s1}
\end{figure}

Based on the GS behavior we selected several representative values of the spin $s$ and the single-ion anisotropy parameter $D/|J|$, for which we present the thermal behavior of the quantities of interest. In particular, we focus on the magnetization, the adiabatic temperature change, and the isothermal entropy change. The respective quantities for $s=1$, representing the integer spin values, and the single-ion anisotropy parameter values $D/|J|=0,-0.5$ and $-1$ are presented in Figure~\ref{fig:ft_T_s1} in the $k_{B}T/|J|-h/|J|$ parameter plane. We note that although here we focus on $D/|J| \leq 0$, for which the GS analysis predicts qualitative changes and rich behavior of the evaluated quantities, later it will be shown that also positive values of $D/|J|$ can significantly influence magnetocaloric properties. The top row shows the evolution of the magnetization normalized by its saturation value, $m/m_{sat}$, with the varying $D/|J|$. The GS sharp magnetization jumps at the critical fields $h_{ci}/|J|$, the number of which increases with the decreasing $D/|J|$, are gradually smeared out at higher $k_{B}T/|J|$. One can also notice that the degree of smearing at (constant) low temperatures increases with the decreasing values of $D/|J|$.

Magnetocaloric properties can be examined by studying the adiabatic temperature changes in the applied magnetic field. Those can be observed by presenting the entropy density in the magnetic field-temperature plane, as shown in the middle row of Figure~\ref{fig:ft_T_s1}. Namely, isentropes represent the temperature response to the variation of the magnetic field under adiabatic conditions. Apparently the most pronounced changes in temperature are displayed in the vicinity of the critical fields at which magnetization jumps occur. In particular, if the magnetic field is set to some critical value and the entropy close to the residual value corresponding to the given critical field, then a dramatic increase (decrease) of temperature can be achieved by a small increase (decrease) of the field above (below) the critical value. Another relevant quantity is the isothermal entropy change, the behavior of which is presented in the bottom row of Figure~\ref{fig:ft_T_s1}. The most dramatic changes are again observed at low temperatures near the critical fields. Namely, for a fixed (low) temperature the field increase (decrease) from $h_{ci}/|J|$ to higher (lower) values is accompanied by a sudden decrease in entropy. However, depending on the entropies at the initial and final field values, $S_i$ and $S_f$, the entropy density change $\Delta S(T,\Delta h)/Nk_B =(S_f-S_i)/Nk_B$ can be either negative or positive. Thus, the system can display either direct or inverse MCE.

\begin{figure}[t!]
\centering
\subfigure{\includegraphics[width=5 cm]{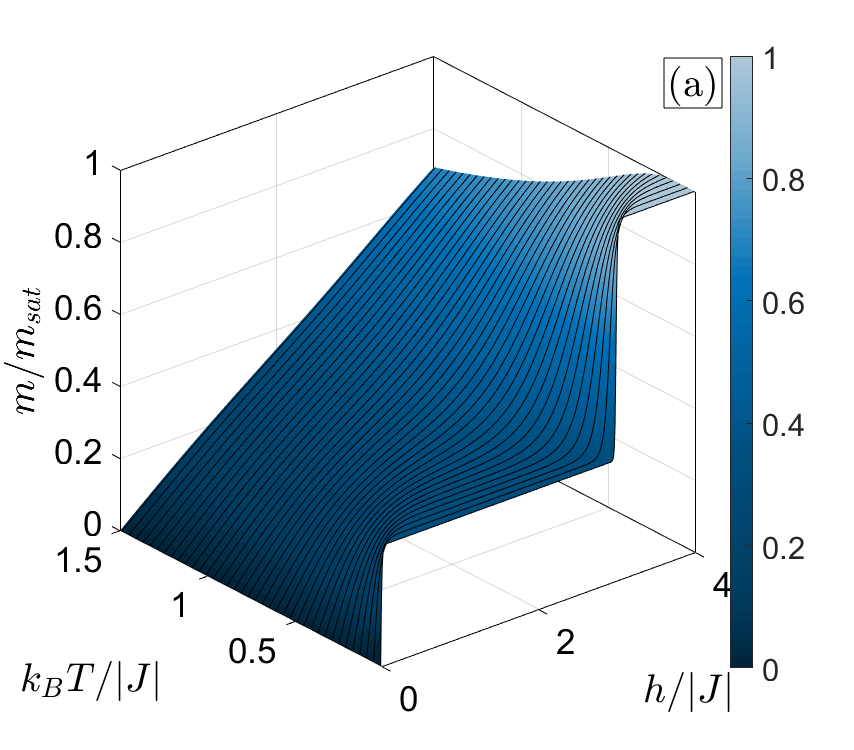}\label{fig:trianglemagn_S1.5_A0}}
\subfigure{\includegraphics[width=5 cm]{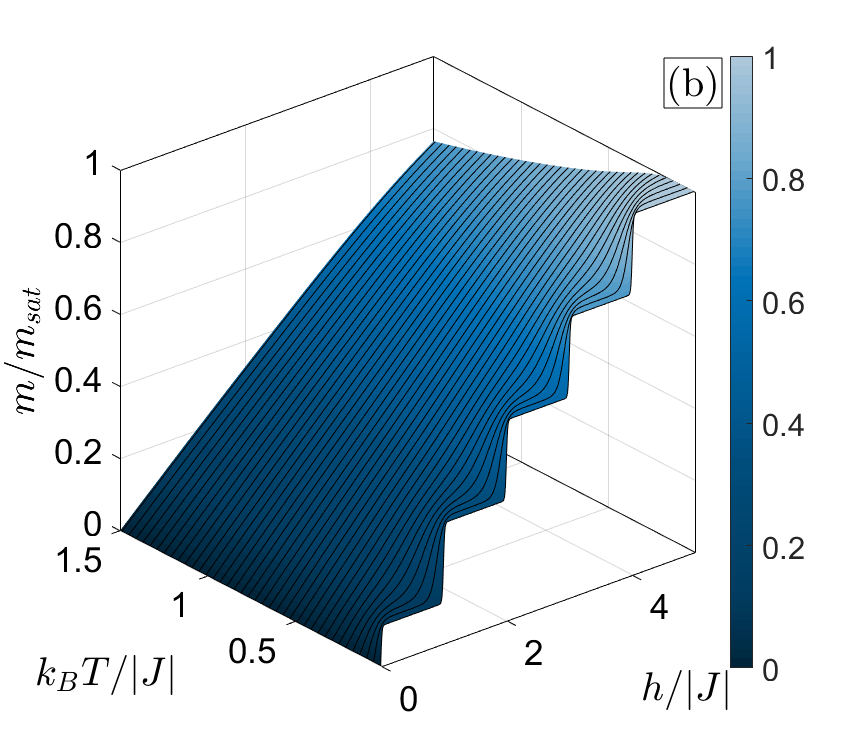}\label{fig:trianglemagn_S1.5_A-0.5}}
\subfigure{\includegraphics[width=5 cm]{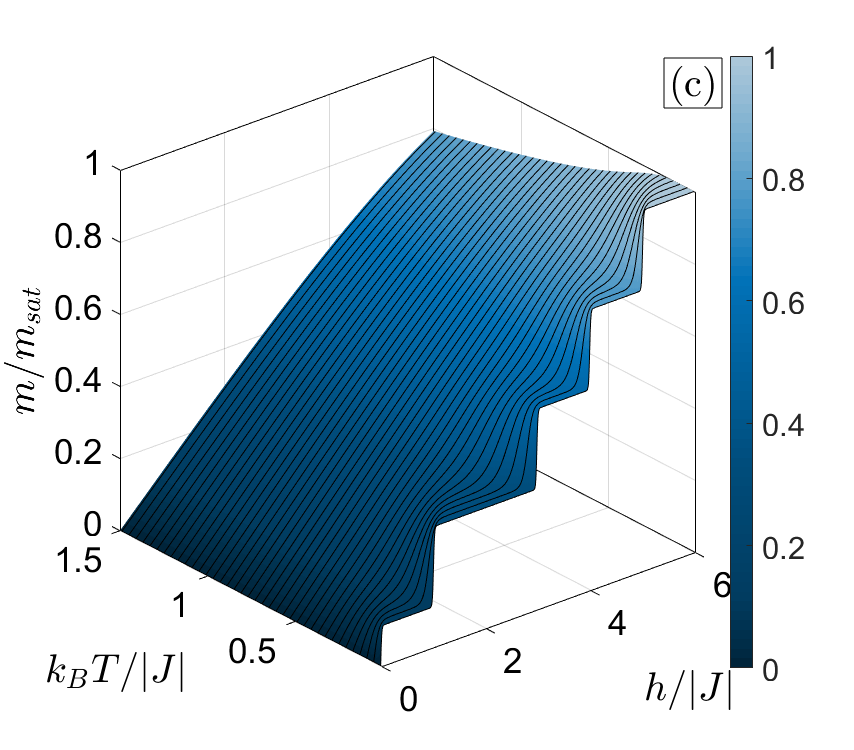}\label{fig:trianglemagn_S1.5_A-1}}\\
\subfigure{\includegraphics[width=5 cm]{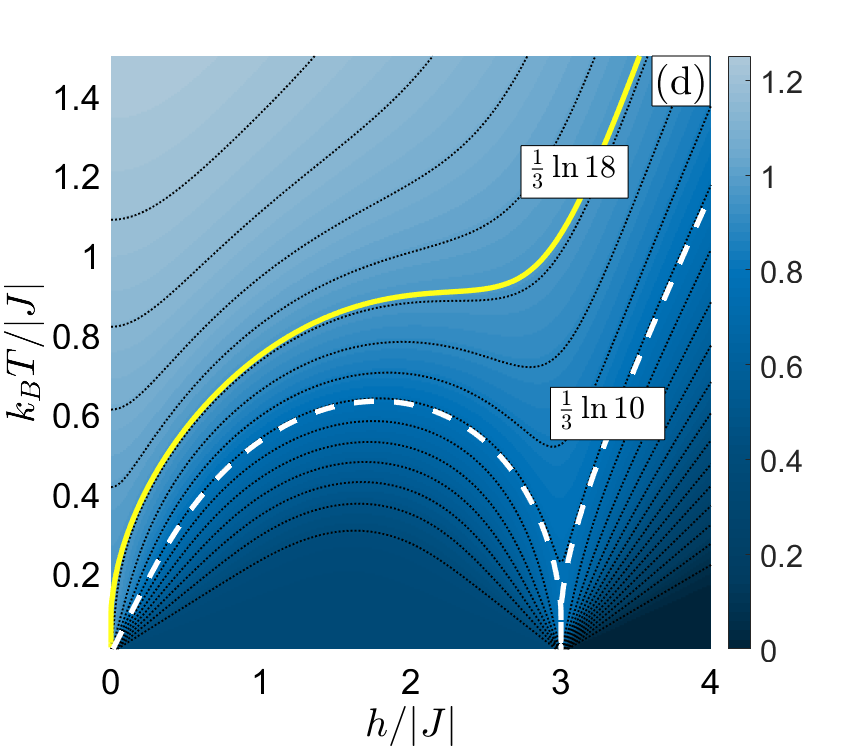}\label{fig:triangleentr_top_S1.5_A0}}
\subfigure{\includegraphics[width=5 cm]{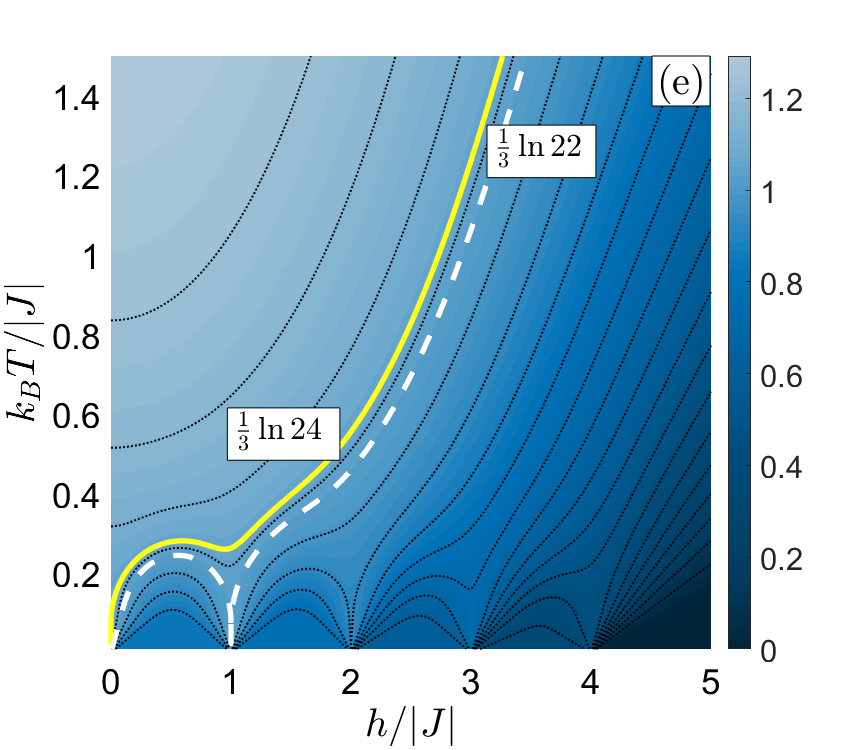}\label{fig:triangleentr_top_S1.5_A-0.5}}
\subfigure{\includegraphics[width=5 cm]{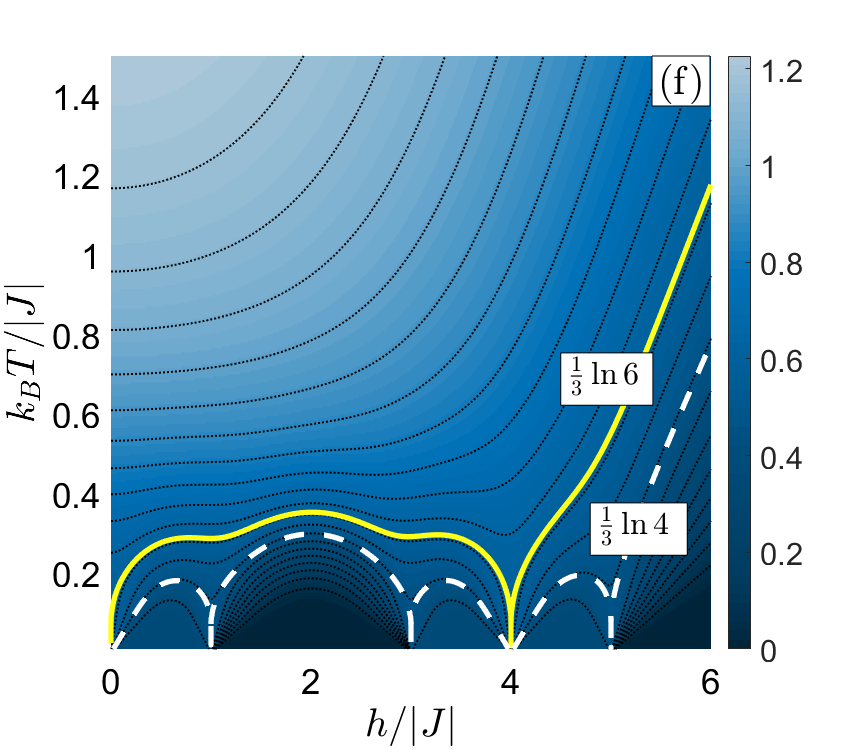}\label{fig:triangleentr_top_S1.5_A-1}}\\
\subfigure{\includegraphics[width=5 cm]{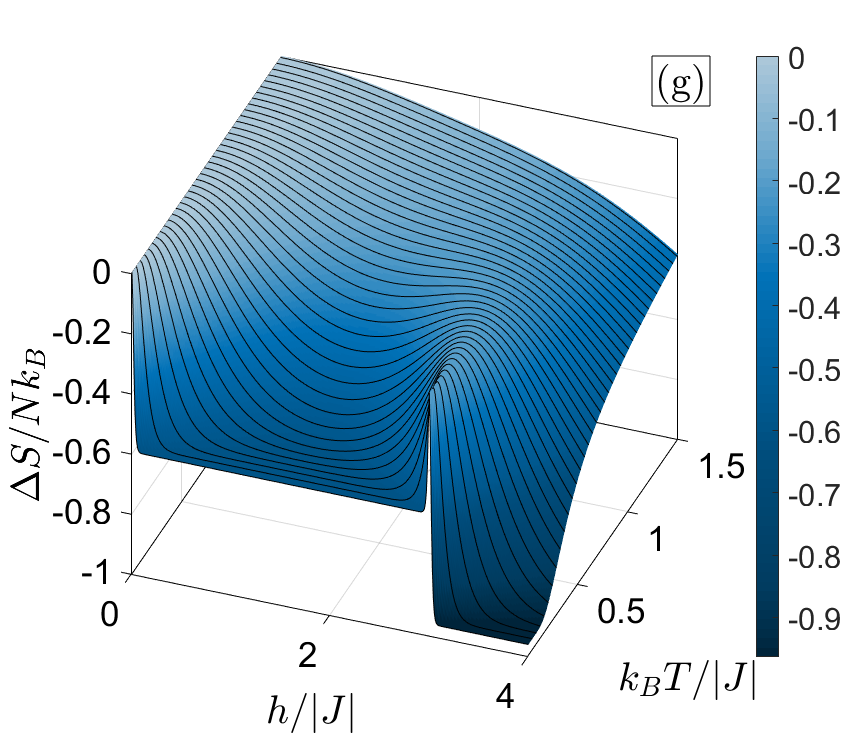}\label{fig:triangleentr_delta_S1.5_A0}}
\subfigure{\includegraphics[width=5 cm]{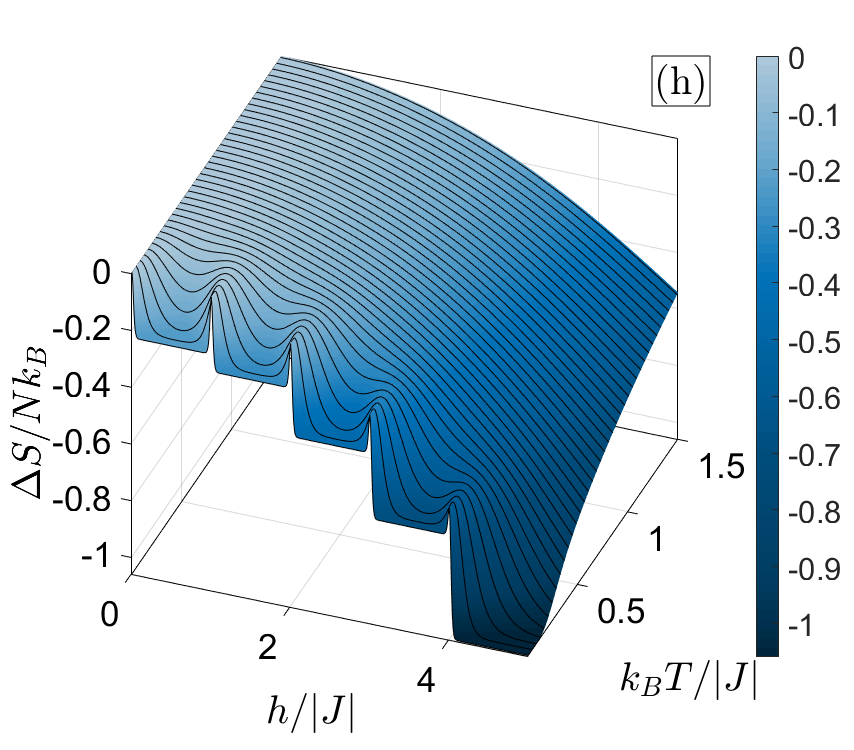}\label{fig:triangleentr_delta_S1.5_A-0.5}}
\subfigure{\includegraphics[width=5 cm]{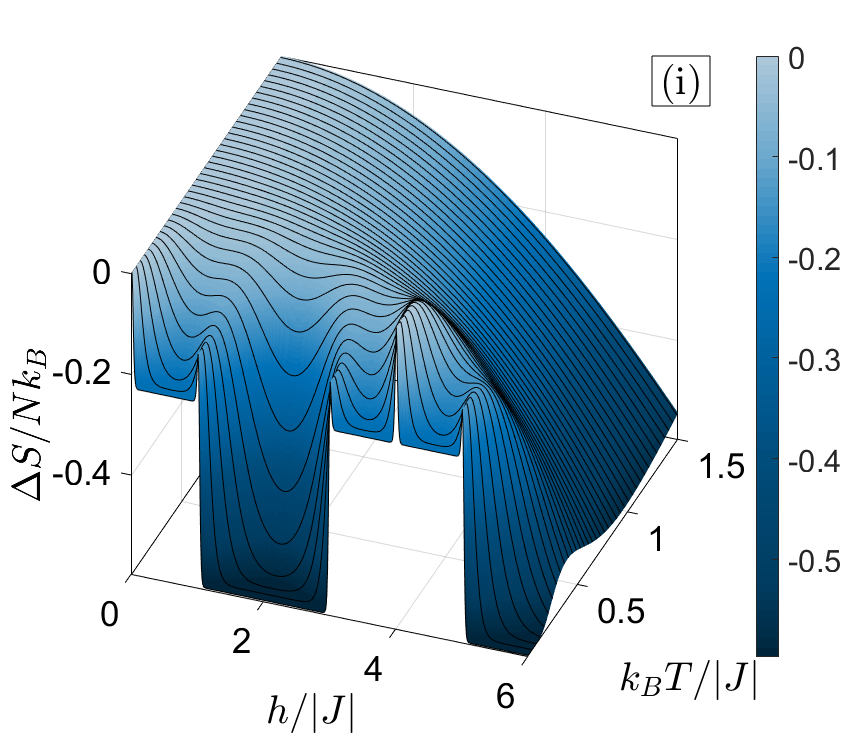}\label{fig:triangleentr_delta_S1.5_A-1}}
\caption{Magnetization (top row), entropy density (middle row), and isothermal entropy density change (bottom row) in $k_{B}T/|J|-h/|J|$ plane, for T cluster with $s=3/2$ and $D/|J|=0$ (first column), $D/|J|=-0.5$ (second column), $D/|J|=-1$ (last column). The highlighted isentropes in the middle row correspond to the GS residual entropies at the first two critical fields $h_{c1}/|J|$ (solid yellow) and $h_{c2}/|J|$ (dashed white).}\label{fig:ft_T_s1.5}
\end{figure}

From practical point of view it is desirable that favorable magnetocaloric properties are displayed within the range of sufficiently small magnetic fields. Even if we restrict ourselves to the first (the smallest) critical field $h_{c1}/|J|$ we still can observe two types of MCE. Namely, for the integer spins and negative $D/|J|$, there is a zero-magnetization plateau and reducing the field below $h_{c1}/|J|>0$ results in increase of temperature under the adiabatic condition and decrease of entropy under the isothermal condition, i.e. the inverse MCE. For the integer spins and non-negative $D/|J|$ (first column in Figure~\ref{fig:ft_T_s1}) as well as the half-integer spins and arbitrary $D/|J|$ (see Figure~\ref{fig:ft_T_s1.5} for the $s=3/2$ case) there is no zero-magnetization plateau and thus $h_{c1}/|J|=0$. Then, by turning off even a very small initial field $h_i$ under the adiabatic condition (adiabatic demagnetization), with the entropy set close to the zero-field residual value, e.g., $S/Nk_B=\ln 12^{1/3}$ in Figure~\ref{fig:triangleentr_top_S1_A0}, one can achieve a rapid cooling of the system. Small fields lift the zero-field degeneracy only partially. Nevertheless, considerable isothermal entropy density changes can be obtained for $-0.5 < D/|J| \leq 0$ and $h_i/|J|<<1$ (Figure~\ref{fig:triangleentr_delta_S1.5_A0}) but also for $D/|J| < -0.5$ and $h_i/|J|\gtrsim 1$ (Figure~\ref{fig:triangleentr_delta_S1.5_A-1}).

\begin{figure}[t!]
\centering
\subfigure{\includegraphics[width=5 cm]{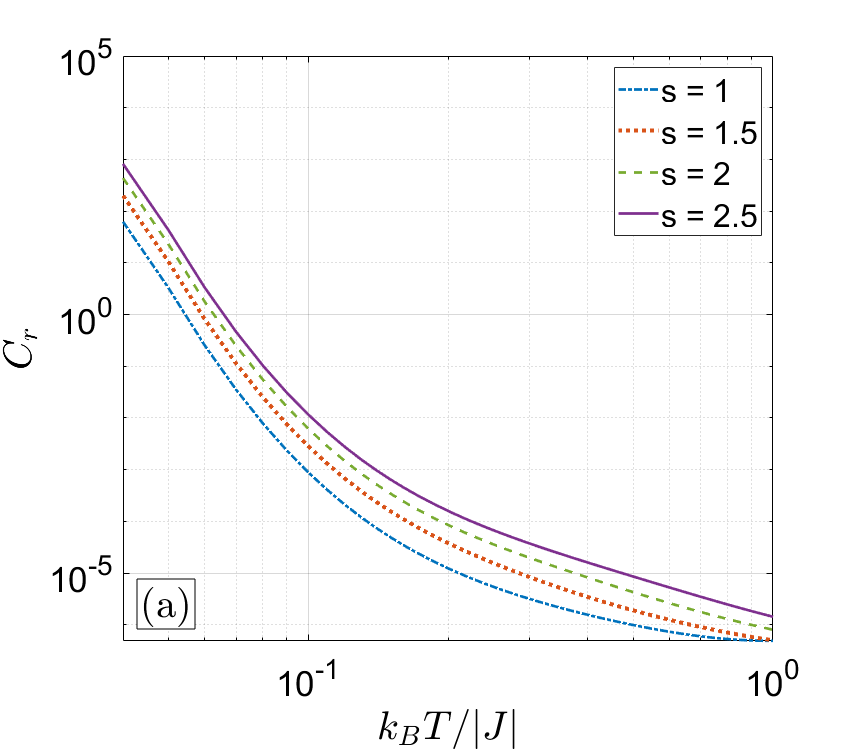}\label{fig:TR_CR_SPDEP_A0}}
\subfigure{\includegraphics[width=5 cm]{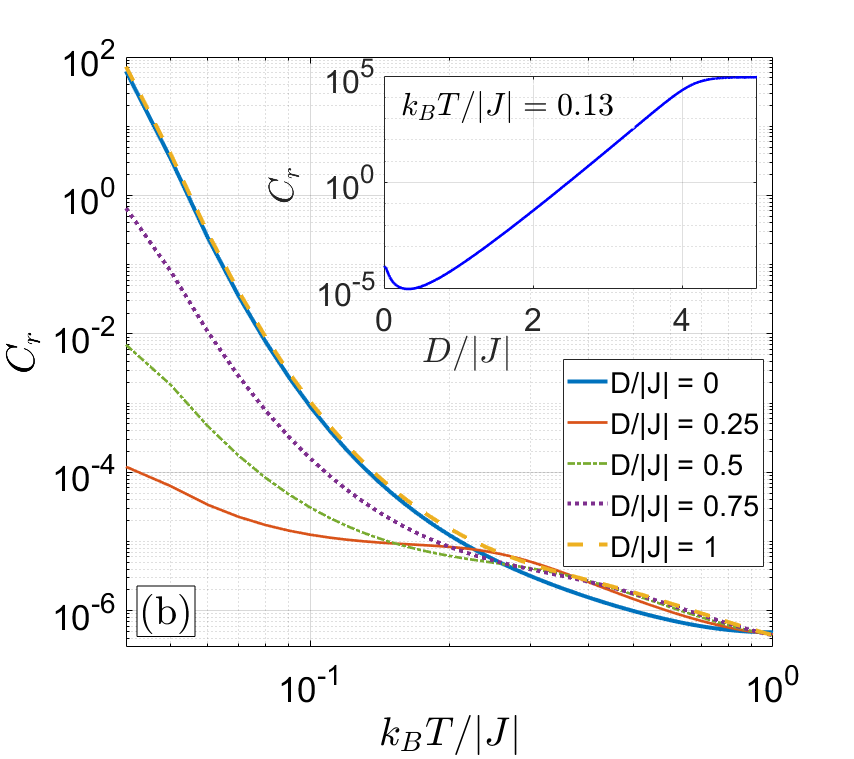}\label{fig:TR_CR_S1}}
\subfigure{\includegraphics[width=5 cm]{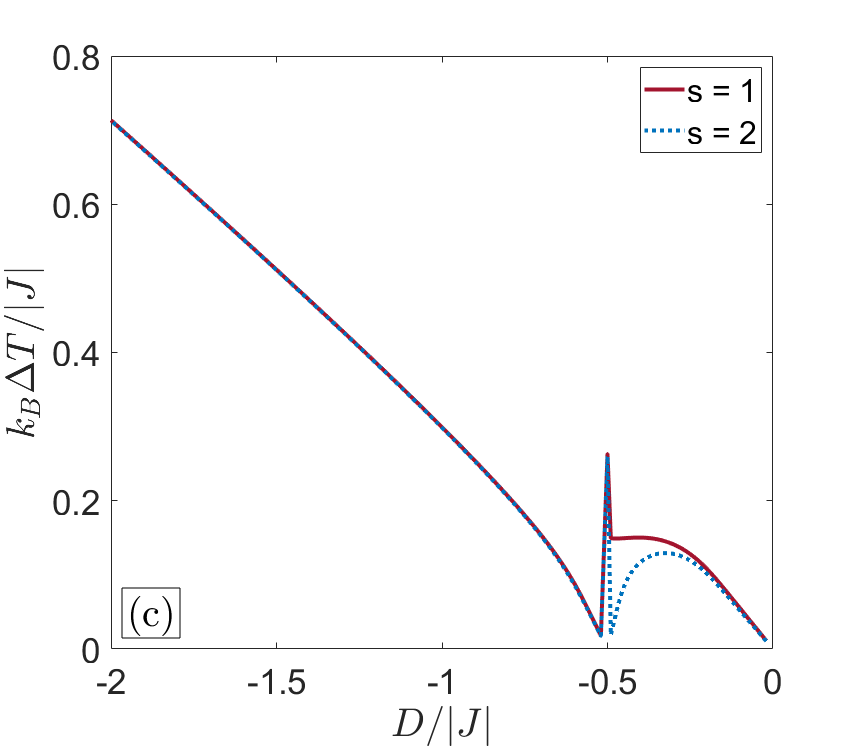}\label{fig:TADI_TR}}
\caption{(a,b) Temperature dependencies of the cooling rates $C_r$ on approach to $h_f=0$ in the adiabatic process with the zero-field GS degeneracy in the T cluster for (a) $D/|J|=0$ and different spin values and (b) for $s=1$ and different single-ion anisotropy. The inset in (b) shows the single-ion anisotropy dependence of $C_r$ for $k_{B}T/|J|=0.13$. (c) Single-ion anisotropy dependence of the temperature change when decreasing the field from $h_i=h_{c1}$ to $h_f=0$ in the adiabatic process with the entropy corresponding to the ground-state degeneracy at $h_{c1}/|J|$.}\label{fig:cr_T}
\end{figure}

Despite qualitative similarities between the behaviors of various quantities for different spins and single-ion anisotropies, there are more or less apparent quantitative differences. As already suggested above, spin values as well as the magnitude of the single-ion anisotropy in the $D/|J|>0$ regime can significantly affect magnetocaloric properties. In the plots shown in the middle rows of Figures~\ref{fig:ft_T_s1} and~\ref{fig:ft_T_s1.5} one can notice that at low fields the isentropes, corresponding to the zero-field GS degeneracies, for $s=3/2$ show noticeably steeper slopes than for $s=1$ and in both cases they dramatically increase in the limit of very low temperatures. In Figure~\ref{fig:TR_CR_SPDEP_A0} we present temperature dependencies of the corresponding cooling rates for different spin values and $D/|J|=0$. Both the increasing spin and decreasing temperature enhance the cooling rate. In the low-temperature region the temperature decrease is accompanied with approximately power-law increase of the cooling rate.

Figure~\ref{fig:TR_CR_S1} shows similar dependence for the fixed spin $s=1$ and increasing positive values of $D/|J|$. It demonstrates that inclusion of a rather small $D/|J|>0$ can lead to appreciable lowering of the cooling rate, especially at very low temperatures. As better shown in the inset for the fixed $k_{B}T/|J|=0.13$, $C_r$ decreases up to $D/|J| \approx 0.33$, at which the cooling rate acquires a minimum value. However, upon further increase of $D/|J|$ it starts increasing and at some value, e.g., $D/|J| \approx 1$ for $s=1$, it surpasses that for $D/|J|=0$. Eventually, at sufficiently large $D/|J|>4$ the cooling rates stabilize (level off) at extremely large values, which are about nine orders of magnitude larger than for $D/|J|=0$. Finally, we demonstrate the effect of negative values of the single-ion anisotropy parameter in the integer spin models, which leads to the appearance of the zero-magnetization plateau, on the inverse MCE. In particular, in Figure~\ref{fig:TADI_TR} we show the single-ion anisotropy dependence of the temperature change when the initial field corresponding to the first critical field value, i.e., $h_i/|J|=h_{c1}/|J|\equiv -D/|J|$, is reduced to $h_f/|J|=0$ in the adiabatic process with the entropy, which corresponds to the ground-state degeneracy at $h_{c1}/|J|$. For $s=1$, the initial steeper increase in $k_{B} \Delta T/|J|$ at $D/|J| \lesssim 0$ is followed by levelling off and a spike-like increase right at $D/|J|=-0.5$. Below $D/|J|=-0.5$ we observe a sharp drop of $k_{B} \Delta T/|J|$ to almost zero and then again initially faster at $D/|J| \lesssim -0.5$ followed by milder almost linear increase with the decreasing $D/|J|$. For $s=2$ the value of $k_{B} \Delta T/|J|$ drops to almost zero as $D/|J|$ approaches the value of $-0.5$ from either side but for $D/|J|<-0.5$ the curves for $s=1$ and $s=2$ collapse on each other and thus no dependence on the spin value is observed.

\subsubsection{2CS cluster}
\begin{figure}[t!]
\centering
\subfigure{\includegraphics[width=5 cm]{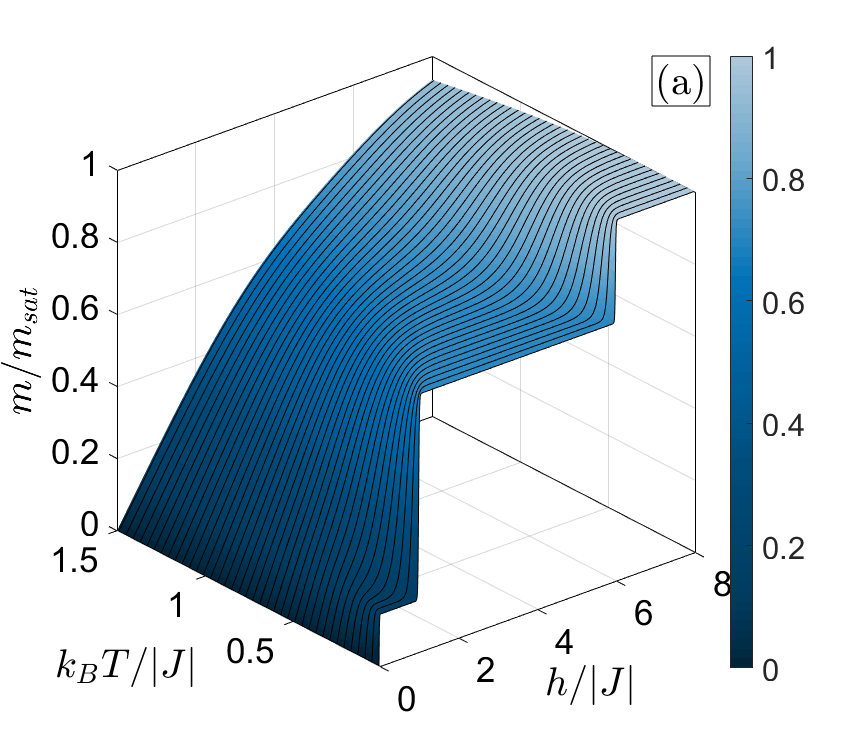}\label{fig:2CSmagn_S1_A0}}
\subfigure{\includegraphics[width=5 cm]{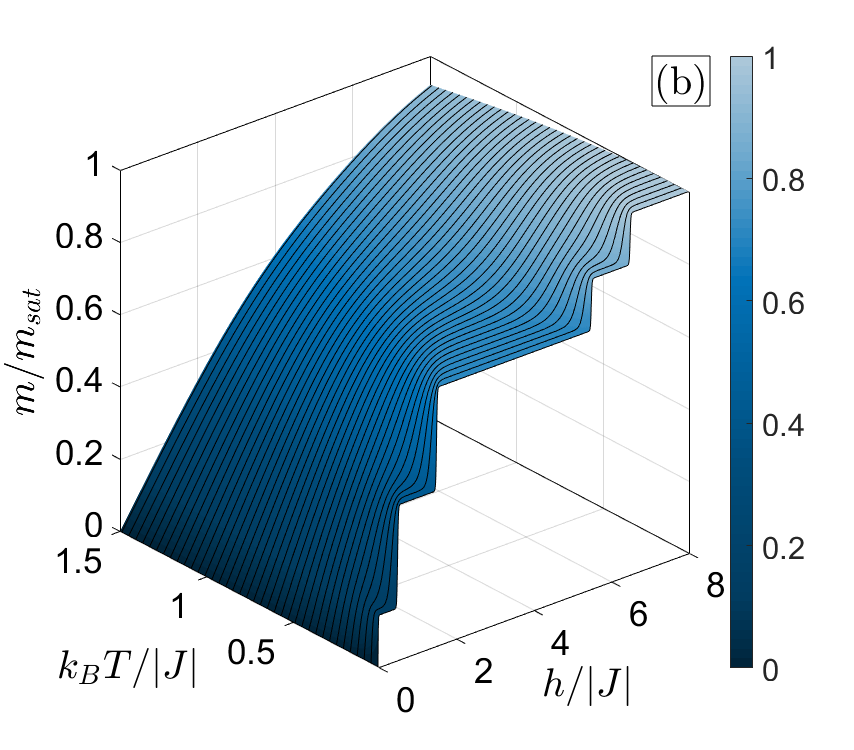}\label{fig:2CSmagn_S1_A-0.5}}
\subfigure{\includegraphics[width=5 cm]{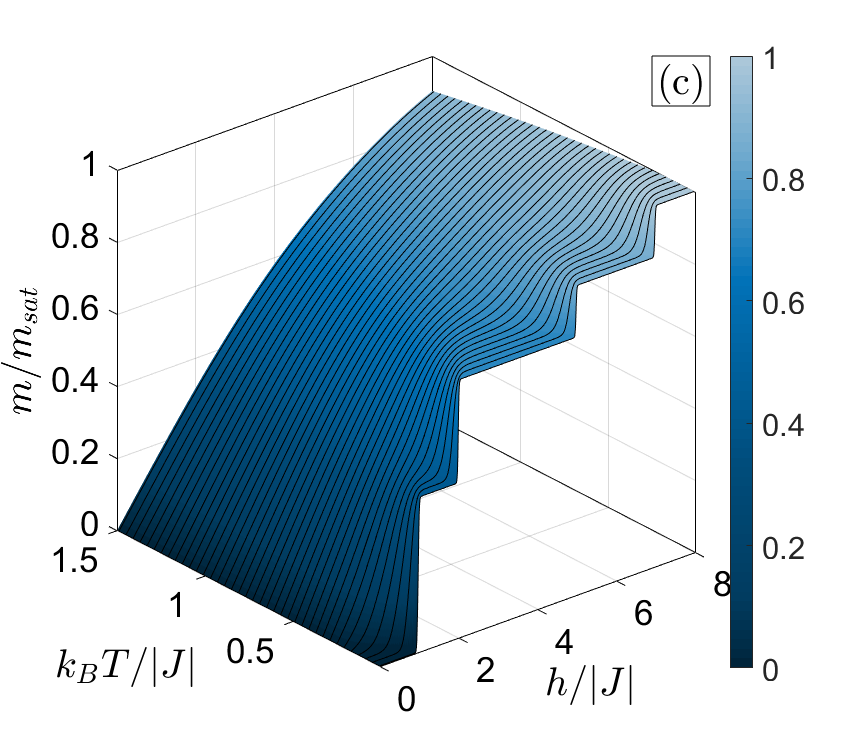}\label{fig:2CSmagn_S1_A-1}}\\
\subfigure{\includegraphics[width=5 cm]{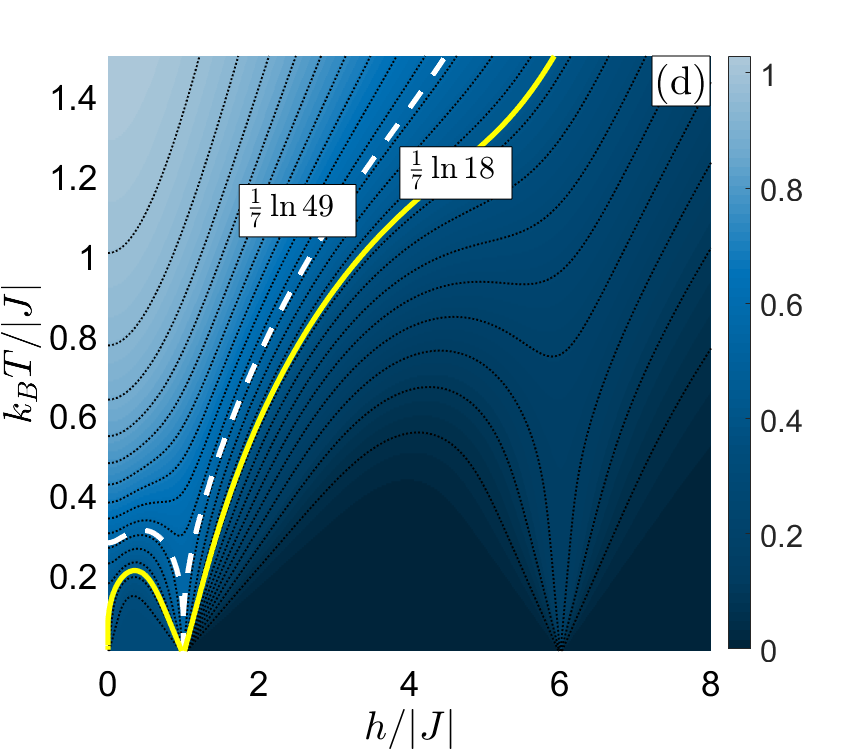}\label{fig:2CSentr_top_S1_A0}}
\subfigure{\includegraphics[width=5 cm]{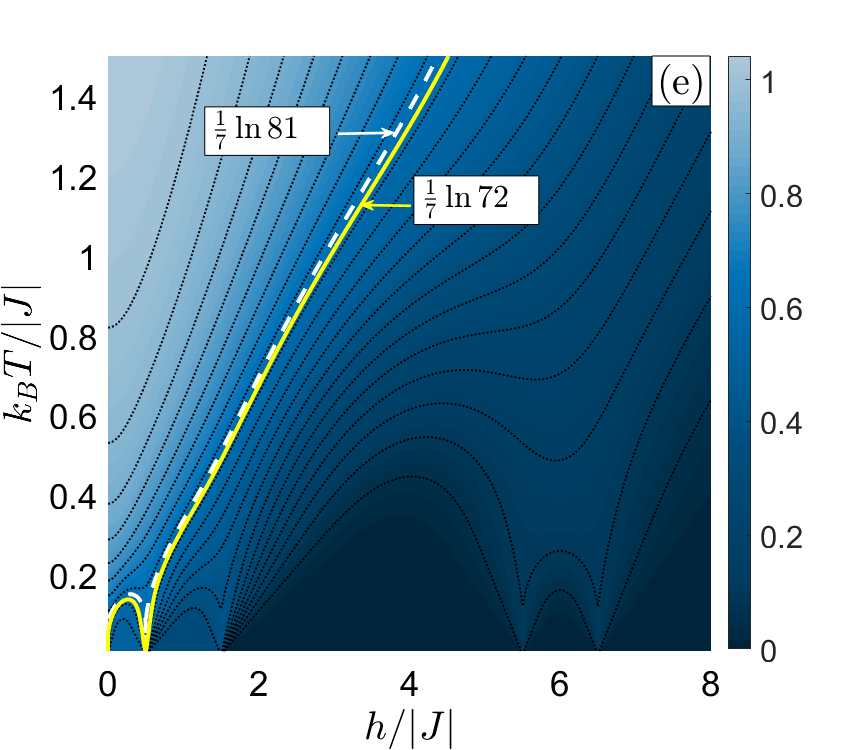}\label{fig:2CSentr_top_S1_A-0.5}}
\subfigure{\includegraphics[width=5 cm]{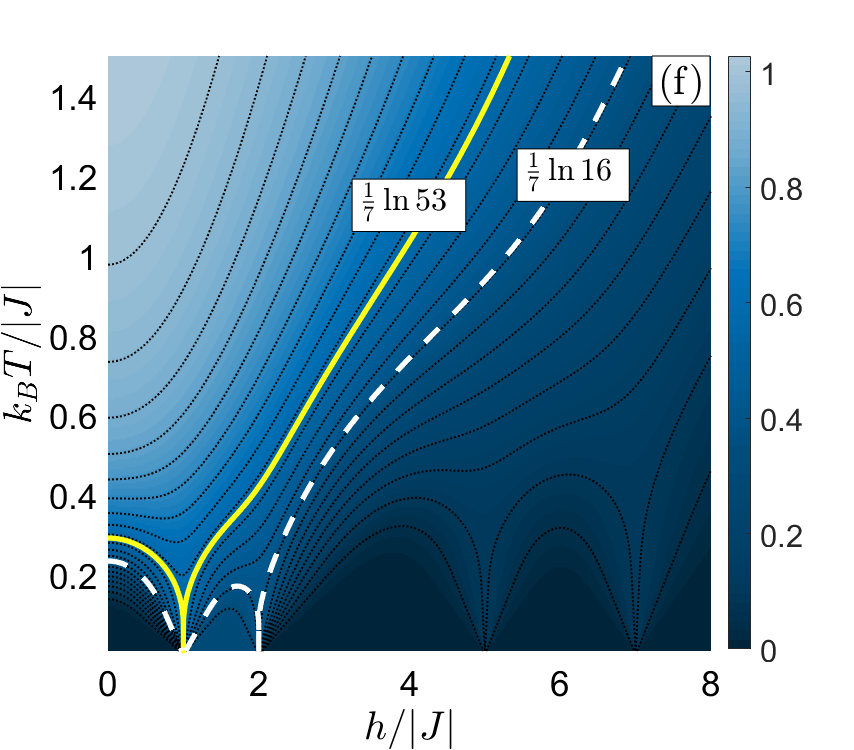}\label{fig:2CSentr_top_S1_A-1}}\\
\subfigure{\includegraphics[width=5 cm]{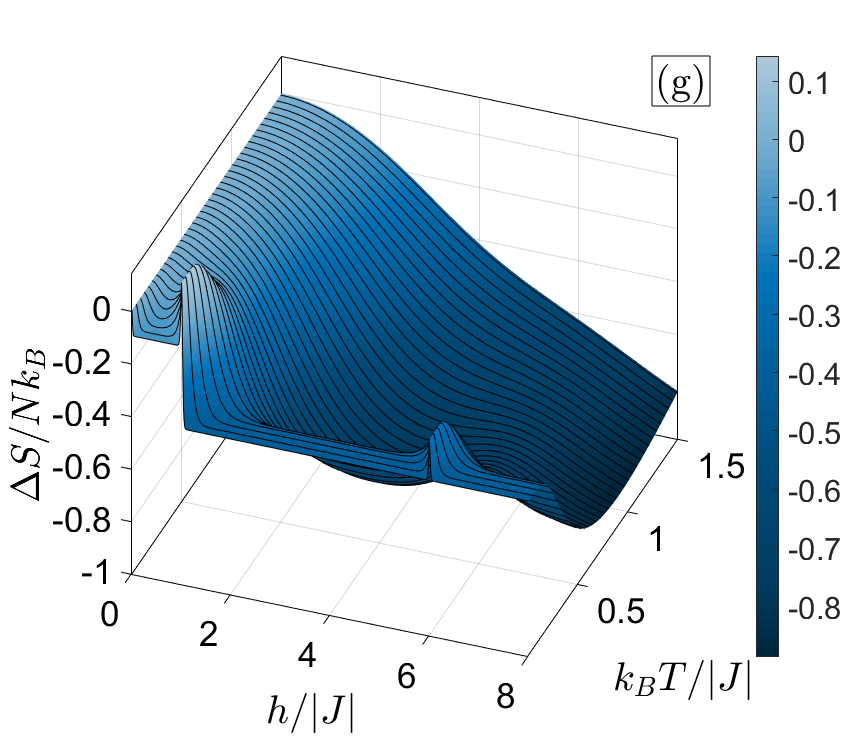}\label{fig:2CSentr_delta_S1_A0}}
\subfigure{\includegraphics[width=5 cm]{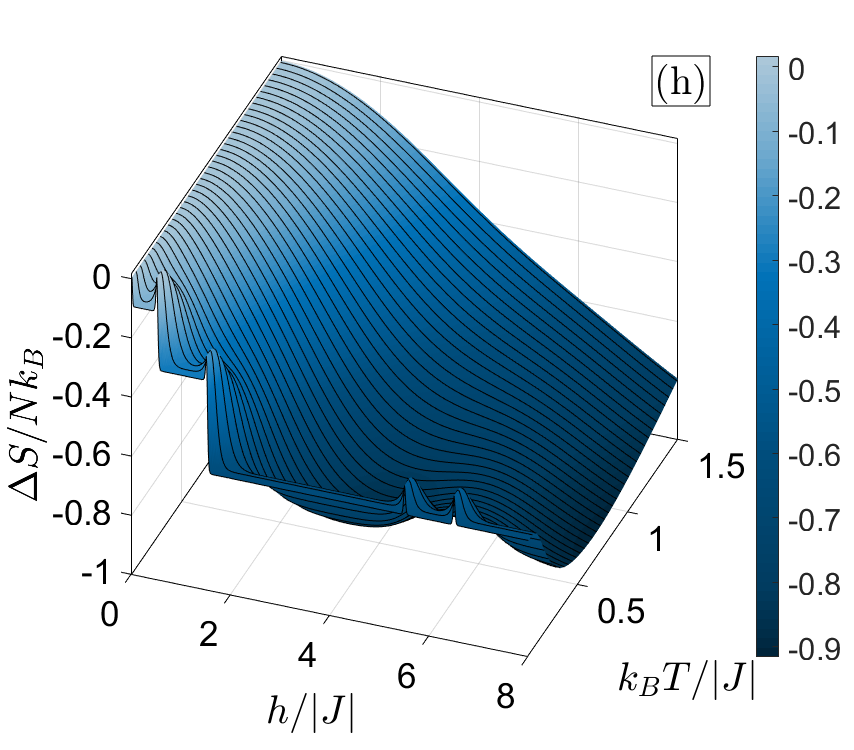}\label{fig:2CSentr_delta_S1_A-0.5}}
\subfigure{\includegraphics[width=5 cm]{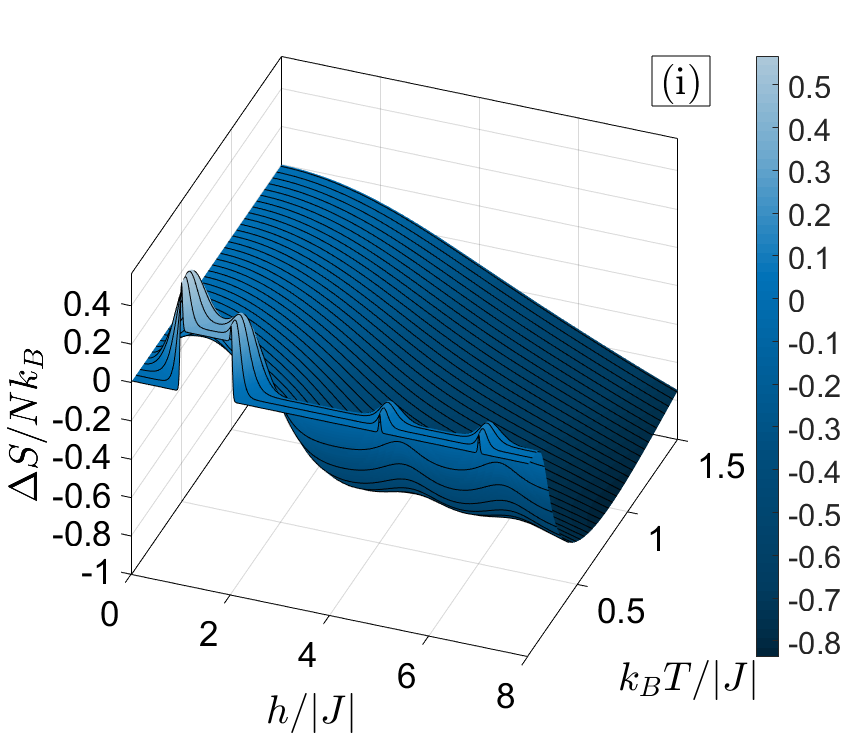}\label{fig:2CSentr_delta_S1_A-1}}
\caption{Magnetization (top row), entropy density (middle row), and isothermal entropy density change (bottom row) in $k_{B}T/|J|-h/|J|$ plane, for 2CS cluster with $s=1$ and $D/|J|=0$ (first column), $D/|J|=-0.5$ (second column), and $D/|J|=-1$ (last column). The highlighted isentropes in the middle row correspond to the GS residual entropies at the first two critical fields $h_{c1}/|J|$ (solid yellow) and $h_{c2}/|J|$ (dashed white).}\label{fig:ft_2CS_s1}
\end{figure}

Finite-temperature analysis of the studied quantities in the 2CS cluster displays some similarities but also some differences compared to the T cluster case. In Figure~\ref{fig:ft_2CS_s1} the magnetization, the adiabatic temperature change, and the isothermal entropy change for $s=1$ and three selected values of the single-ion anisotropy parameter $D/|J|$ are shown in the $k_{B}T/|J|-h/|J|$ parameter plane. Like in the T cluster, for $D/|J| \geq 0$ there is no zero-magnetization plateau, i.e. $h_{c1}/|J|=0$, but instead of one there are two intermediate plateaus with the second critical field $h_{c2}/|J|=s$ (i.e., half of that in the T cluster) at which the magnetization jumps from $1/3$ to $2/3$ of the saturation value but beyond which the system degeneracy is completely removed, except for a single point at $h/|J|=6s$. Thus, the adiabatic demagnetization performed in very small fields with the entropy close to the zero-field residual value leads to a fast cooling to extremely low temperatures also in the 2CS system. It is interesting to notice that for $D/|J|=0$ the GS degeneracy at $h_{c2}/|J|$ is larger that that at $h_{c1}/|J|$ and thus one can achieve the inverse MCE if the field is reduced from $h_i=h_{c2}$ to $h_f=h_{c1}$ (Figure~\ref{fig:2CSentr_delta_S1_A0}). Similar to the T cluster, in the negative $D/|J|$ region the number of the magnetization plateaus is increased. The zero-height plateau is also observed in the integer spin models, however, only if $D/|J|<-0.66$ for $s=1$ and $D/|J|<-0.77$ for $s=2$, leading to the inverse MCE. 

\begin{figure}[t!]
\centering
\subfigure{\includegraphics[width=5 cm]{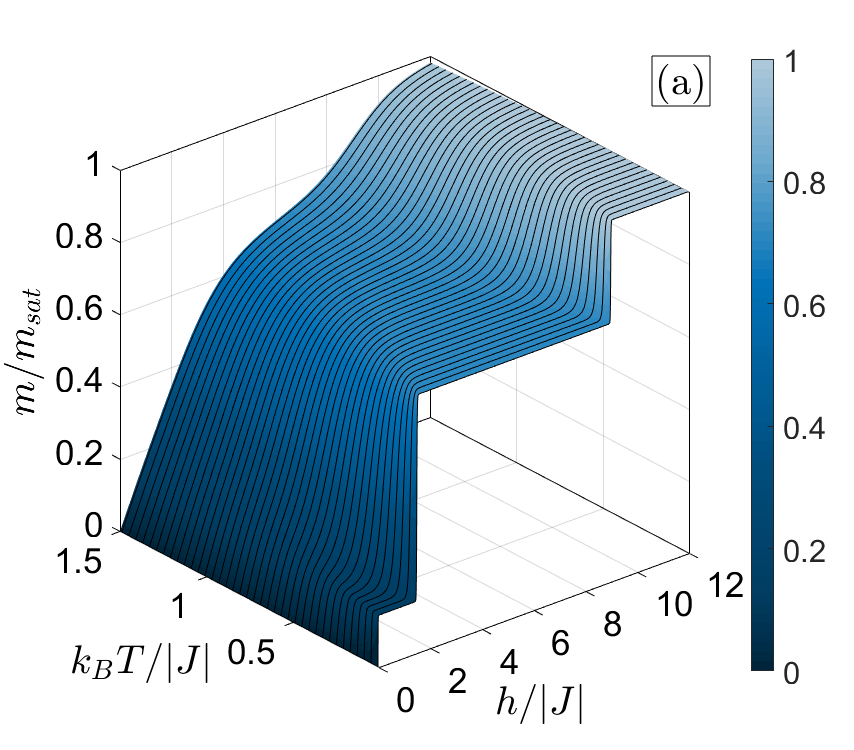}\label{fig:2CSmagn_S1.5_A0}}
\subfigure{\includegraphics[width=5 cm]{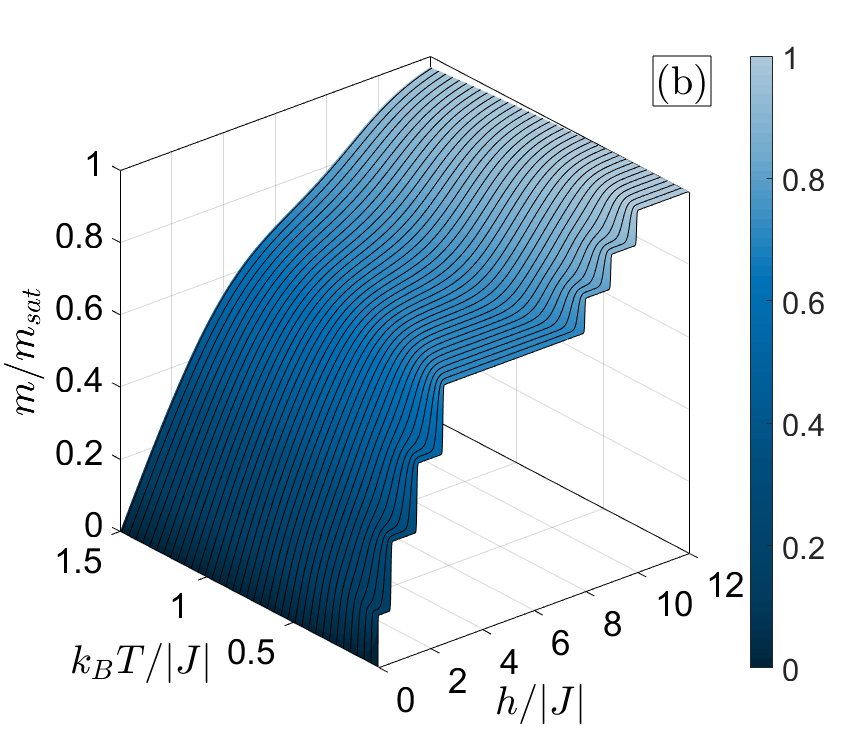}\label{fig:2CSmagn_S1.5_A-0.5}}
\subfigure{\includegraphics[width=5 cm]{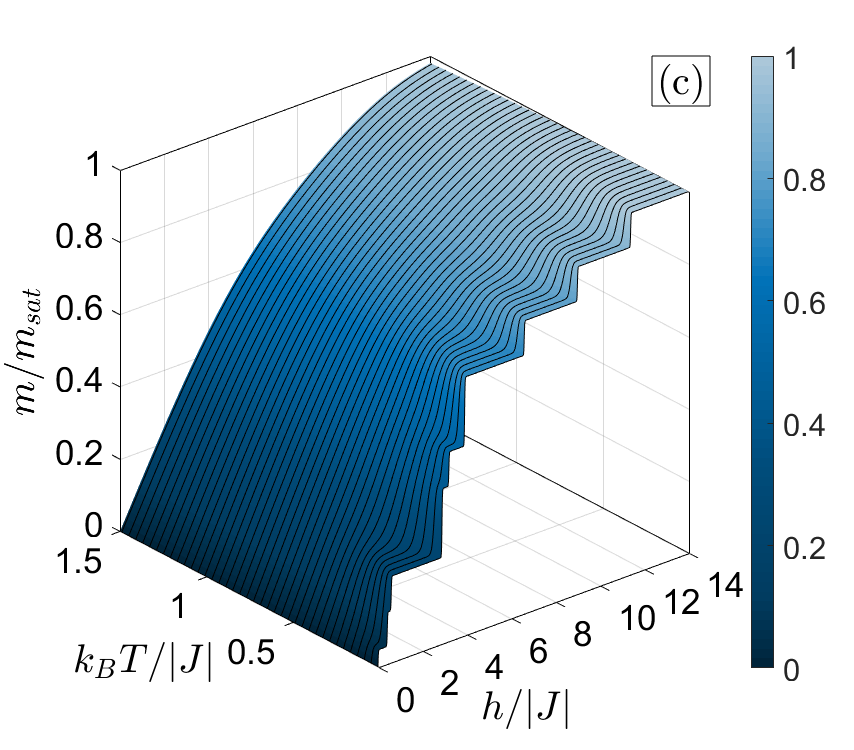}\label{fig:2CSmagn_S1.5_A-1.2}}\\
\subfigure{\includegraphics[width=5 cm]{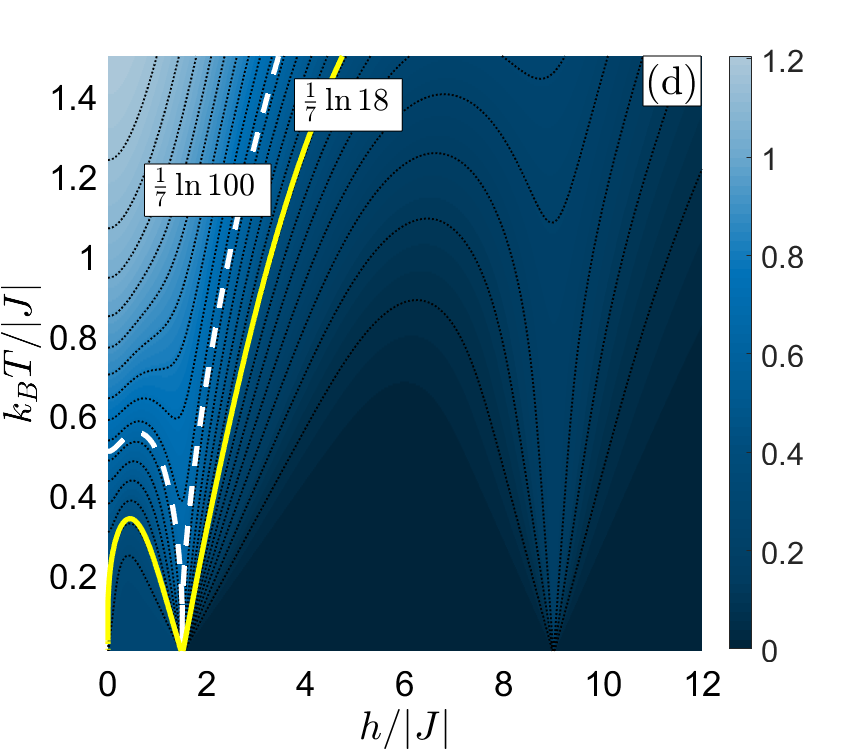}\label{fig:2CSentr_top_S1.5_A0}}
\subfigure{\includegraphics[width=5 cm]{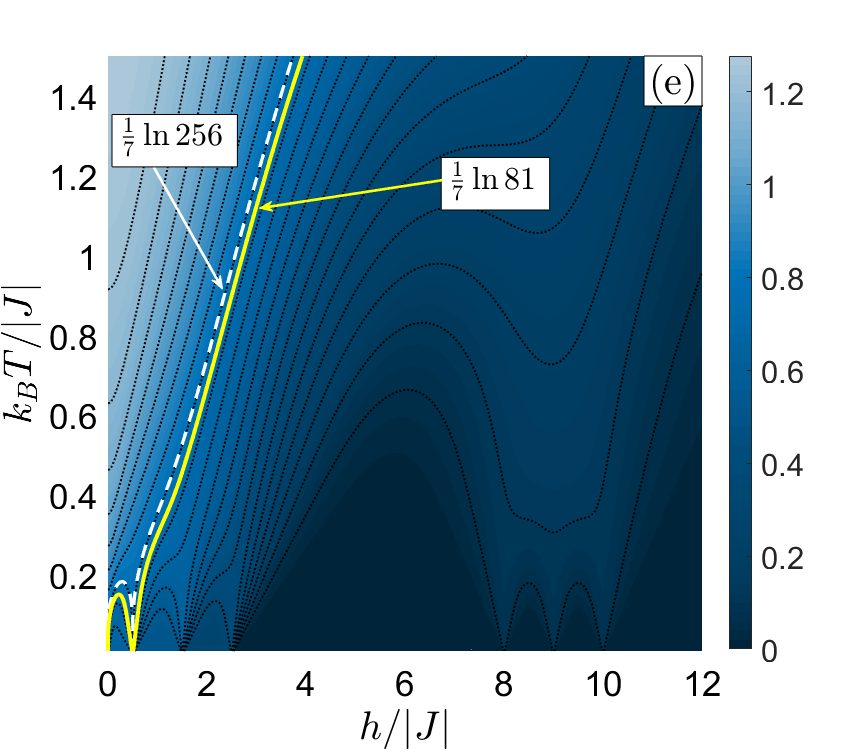}\label{fig:2CSentr_top_S1.5_A-0.5}}
\subfigure{\includegraphics[width=5 cm]{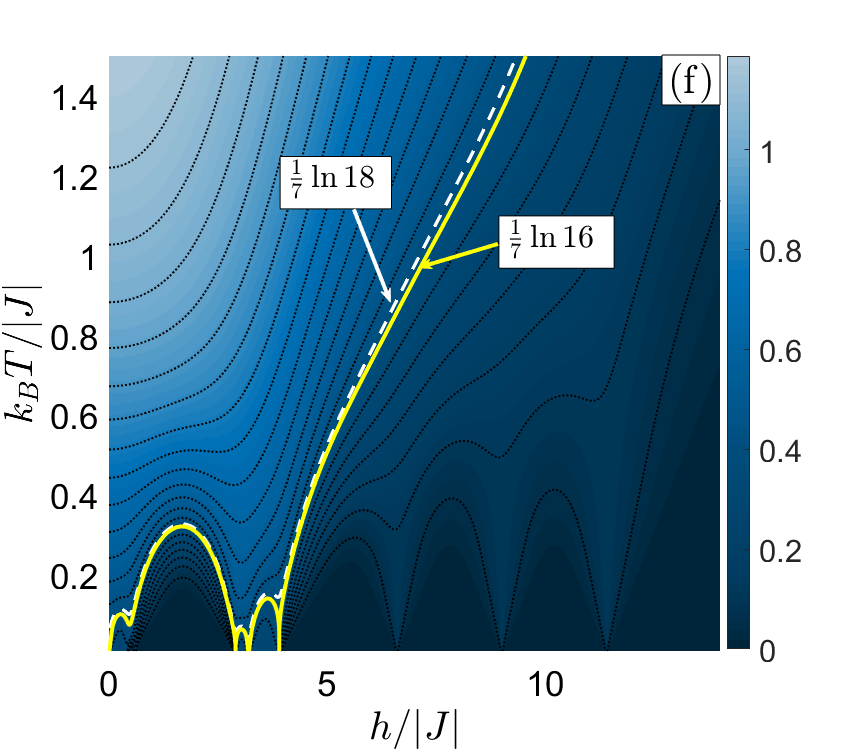}\label{fig:2CSentr_top_S1.5_A-1.2}}\\
\subfigure{\includegraphics[width=5 cm]{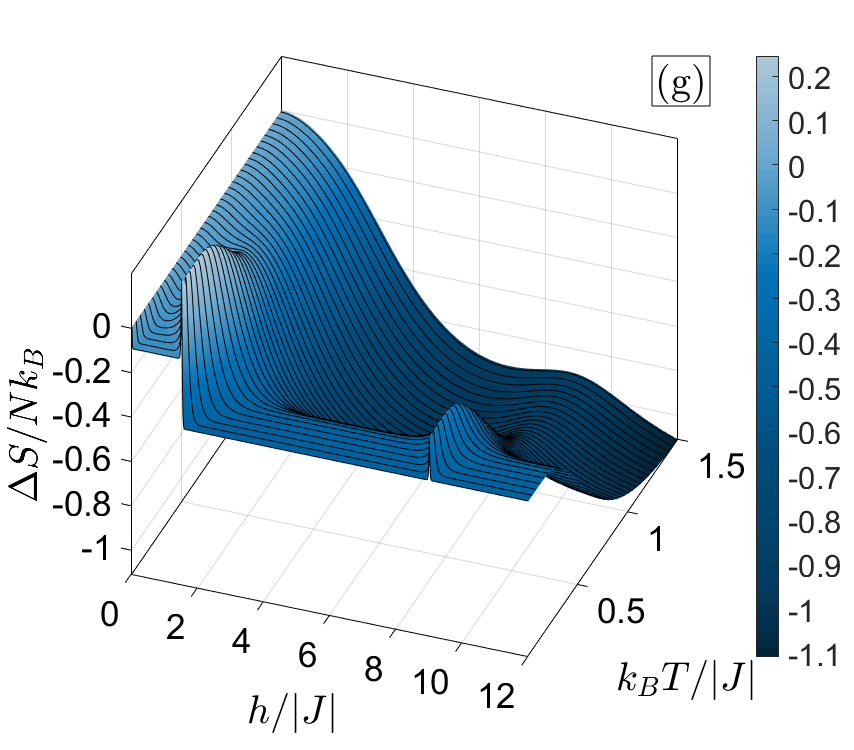}\label{fig:2CSentr_delta_S1.5_A0}}
\subfigure{\includegraphics[width=5 cm]{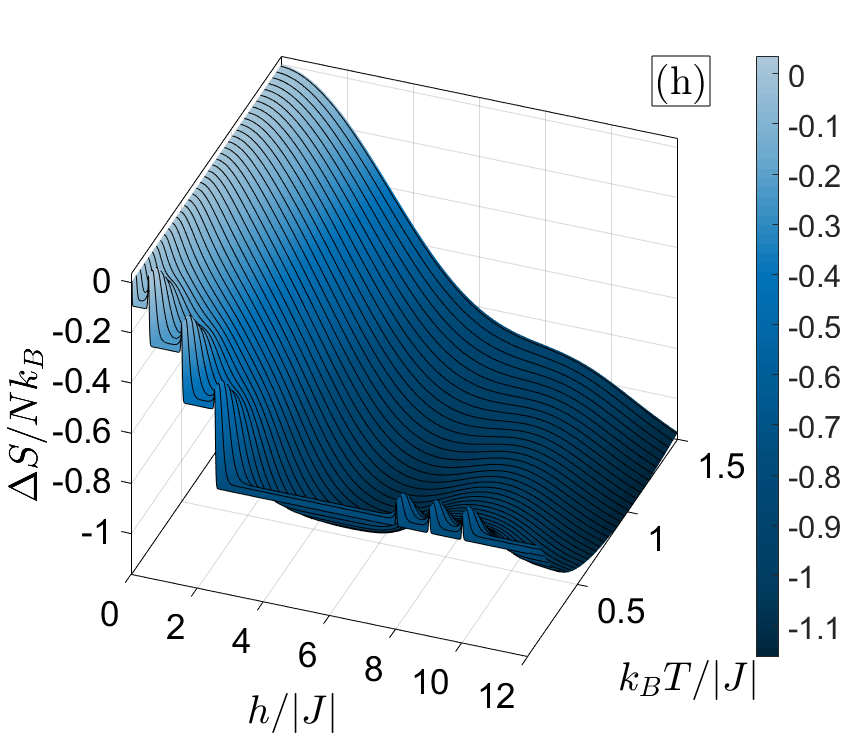}\label{fig:2CSentr_delta_S1.5_A-0.5}}
\subfigure{\includegraphics[width=5 cm]{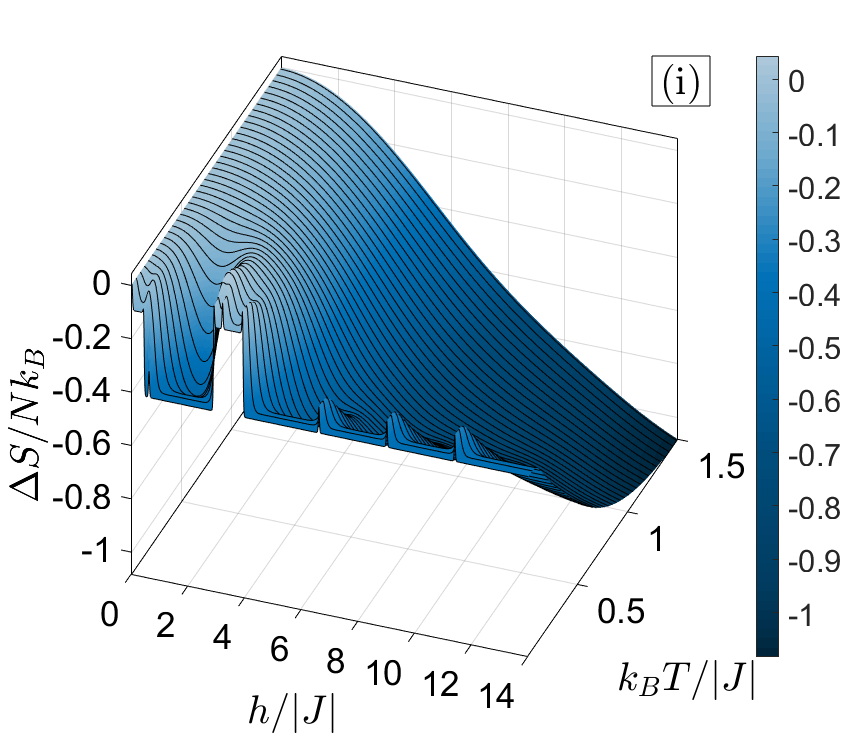}\label{fig:2CSentr_delta_S1.5_A-1.2}}
\caption{Magnetization (top row), entropy density (middle row), and isothermal entropy density change (bottom row) in $k_{B}T/|J|-h/|J|$ plane, for 2CS cluster with $s=3/2$ and $D/|J|=0$ (first column), $D/|J|=-0.5$ (second column), and $D/|J|=-1.2$ (last column). The highlighted isentropes in the middle row correspond to the GS residual entropies at the first two critical fields $h_{c1}/|J|$ (solid yellow) and $h_{c2}/|J|$ (dashed white).}\label{fig:ft_2CS_s1.5}
\end{figure}

The situation for the half-integer spin models is presented in Figure~\ref{fig:ft_2CS_s1.5} by showing similar dependencies for $s=3/2$. In these models the zero-magnetization plateau is absent and the enhanced MCE can be detected for any $D/|J|$ during the adiabatic demagnetization of the system in a very low field region, assuming that the entropy density is close enough to the ground-state degeneracy at zero field. Similarly to the $s=1$ case, the GS degeneracy at $h_{c2}/|J|$ is larger than that at $h_{c1}/|J|$ leading to the inverse MCE if the field is reduced from $h_i=h_{c2}$ to $h_f=h_{c1}$ but not only for the isotropic case but for any $-0.5 < D/|J| \leq 0$ (Figures~\ref{fig:2CSentr_delta_S1.5_A0} and~\ref{fig:2CSentr_delta_S1.5_A-0.5}). As shown in Figure~\ref{fig:2CSentr_delta_S1.5_A-0.5}, for $D/|J|=-0.5$ relatively large isothermal entropy changes can be achieved by decreasing the field from moderate values of $h_i/|J| \gtrsim 2.5$ to zero. However, this is the case only at exactly $D/|J|=-0.5$, which in practice might be difficult to achieve. On the other hand, for $D/|J|<-1$ there is a stable region where somewhat smaller isothermal entropy changes can be achieved but for much smaller $h_i \gtrsim 0.5$ (see Figure~\ref{fig:2CS_entr_1.5} above for the situation at $T=0$).

\begin{figure}[t!]
\centering
\subfigure{\includegraphics[width=5 cm]{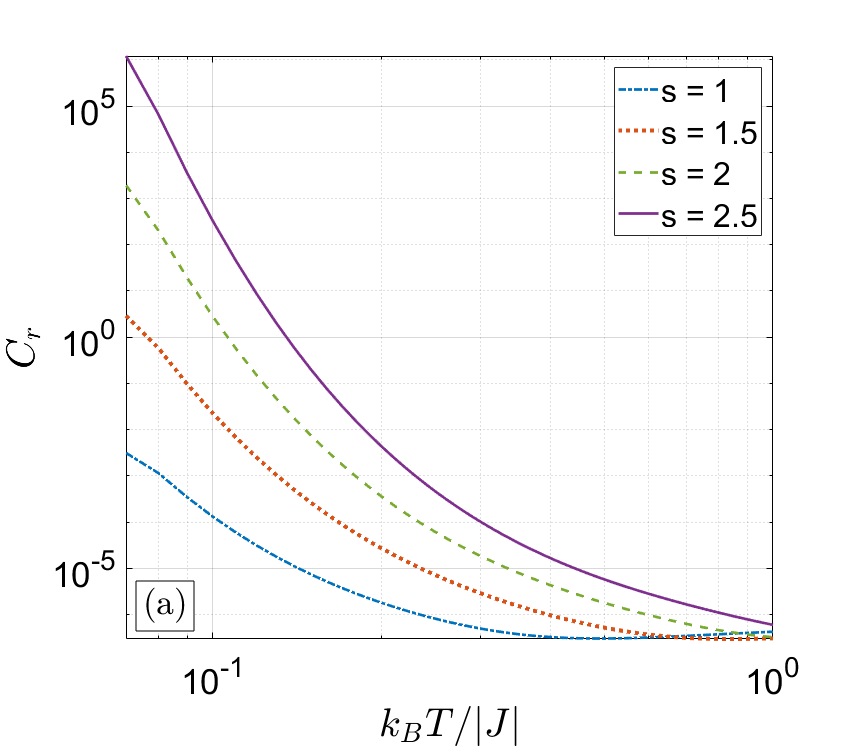}\label{fig:2CS_CR_SPDEP_A0}}
\subfigure{\includegraphics[width=5 cm]{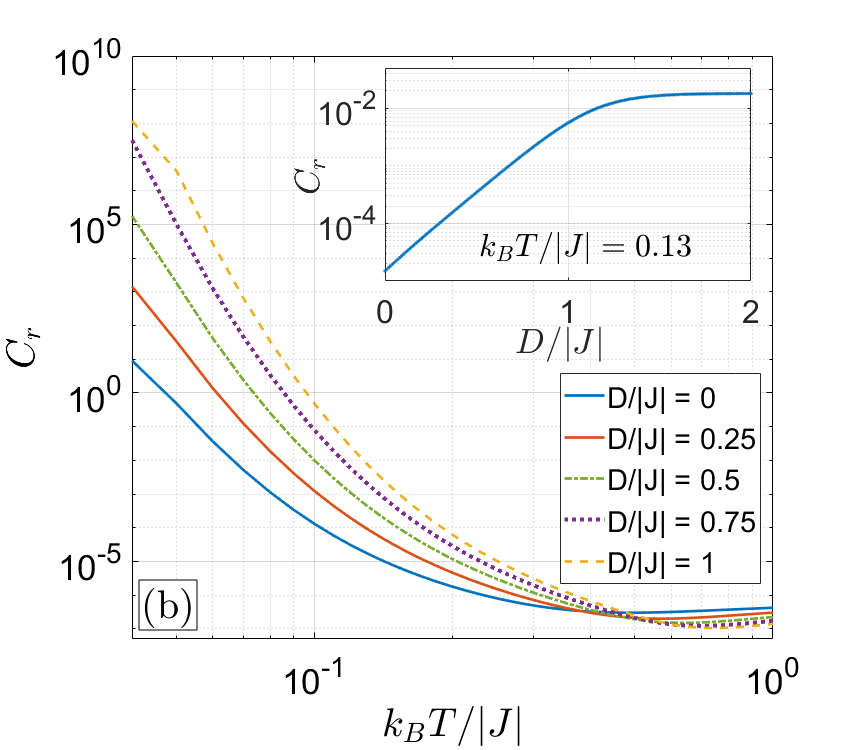}\label{fig:2CS_CR_S1}}
\subfigure{\includegraphics[width=5 cm]{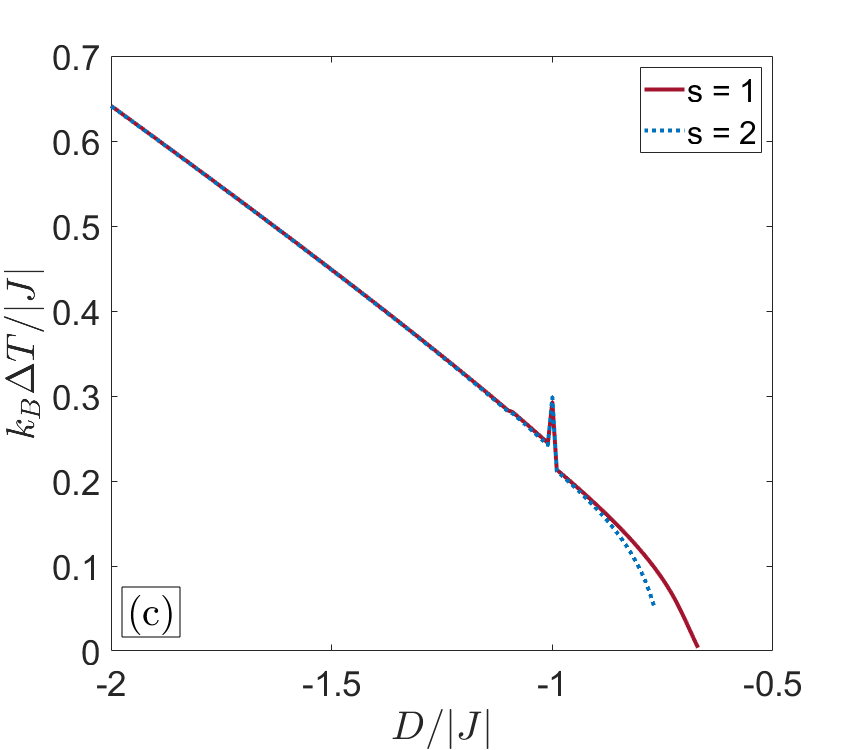}\label{fig:TADI_2CS}}
\caption{(a,b) Temperature dependencies of the cooling rates $C_r$ on approach to $h_f=0$ in the adiabatic process with the zero-field GS degeneracy in the 2CS cluster for (a) $D/|J|=0$ and different spin values and (b) for $s=1$ and different single-ion anisotropy. The inset in (b) shows the single-ion anisotropy dependence of $C_r$ for $k_{B}T/|J|=0.13$. (c) Single-ion anisotropy dependence of the temperature change when decreasing the initial field from $h_i=h_{c1}$ to $h_f=0$ in the adiabatic process with the entropy corresponding to the ground-state degeneracy at $h_{c1}/|J|$.}\label{fig:cr_2CS}
\end{figure}

Temperature dependencies of the cooling rates in the 2CS cluster for different spin values and $D/|J|=0$ are shown in Figure~\ref{fig:2CS_CR_SPDEP_A0}. Like in the T cluster the cooling rate increases with both the increasing spin and decreasing temperature, however, the impact of the spin value on the cooling rate in the 2CS cluster is significantly larger. In the low-temperature region even the smallest increase in the spin value results in a dramatic increase of $C_r$ by several orders. Thus, for example at $k_{B}T/|J|=0.07$ in the T cluster $C_r \approx 10^{-1}$ for any $1<s<5/2$, in the 2CS cluster it ranges between $\approx 10^{-3}$ for $s=1$ and $\approx 10^{6}$ for $s=5/2$. In contrast to the T cluster, the single-ion anisotropy dependence of the cooling rate in the 2CS cluster is monotonically increasing with the levelling off at some value, which is much lower than in the T cluster but is achieved at much smaller $D/|J|$ (Figure~\ref{fig:2CS_CR_S1}). For the integer spins and $D/|J|<-0.66$ for $s=1$ and $D/|J|<-0.77$ for $s=2$, the single-ion anisotropy dependence of the temperature change in the adiabatic process of decreasing the initial field from $h_i=h_{c1}$ to $h_f=0$ with the entropy corresponding to the ground-state degeneracy at $h_{c1}/|J|$ shows similar behavior (Figure~\ref{fig:TADI_2CS}). As $D/|J|$ is decreased the initial steeper increase in $k_{B} \Delta T/|J|$ is followed by a more gentle almost linear increase with a spike-like anomaly at $D/|J|=-1$. 

\section{Discussion}
We studied effects of the single-ion anisotropy on the ground-state and thermodynamic properties in two selected spin-$s$ antiferromagnetic Ising spin clusters with geometrical frustration focusing on MCE. In line with some previous studies~\cite{efre06,efre08,huch11}, inclusion of the single-ion anisotropy in small antiferromagnetic spin clusters with $s\geq 1$ can result in a much more complex energy level crossing behavior. The latter is characteristic for negative values of the single-ion anisotropy parameter $D/|J|$. With increasing $s$ the ground-state magnetization as a function of the external field gradually shows increasing number of plateaus of various heights. For the integer values of $s$ the first plateau is of zero height for any $D/|J|<0$ in the case of the T and for $D/|J| < D_0/|J| \leq 0$, where $D_0$ depends on $s$, in the case of the 2CS cluster.

The presence or absence of the zero-magnetization plateau is crucial for defining the character of the low-temperature MCE in the adiabatic demagnetization process. In the latter case it can lead to a rapid decrease (with up to an infinite gradient) in temperature as the magnetic field vanishes to zero, which can be technologically used for efficient refrigeration to ultra-low temperatures. On the other hand, if the zero-magnetization plateau is present, the temperature increases in the adiabatic demagnetization process at low-temperatures and fields. The two types of the spin clusters show qualitatively similar behavior. However, there are considerable quantitative differences, which have a significant impact on their magnetocaloric properties. Namely, in the isotropic case of $D/|J| = 0$ we observed that the cooling rate $C_r$ in the adiabatic demagnetization process increases with the spin value $s$ in both clusters, however, in the 2CS cluster the increase is by several orders of magnitude faster than in the T cluster. Inclusion of the positive anisotropy in the T cluster at low temperatures results in the initial drop of $C_r$ for small enough $D/|J|$, followed by the approximately exponential increase over several orders of magnitude and eventual levelling-off for sufficiently large $D/|J|$. On the other hand, the increase of $C_r$ in the 2CS cluster is monotonic and the levelling-off occurs at $D/|J|$ much smaller than in the T cluster. For the integer $s$ in the negative $D/|J|$ regime, corresponding to the presence of the zero-magnetization plateau, with the decrease of $D/|J|$ one can observe an almost linear increase of temperature in the adiabatic process of reducing the field to zero with the entropy corresponding to the ground-state degeneracy at the first critical field (magnetization jump) in both clusters.

In the light of the previous studies of the present spin systems with $s=1/2$~\cite{zuko15,mohy19}, which showed favorable magnetocaloric properties, one can conclude that by increasing the spin value and by a proper tuning of the single-ion anisotropy a further significant improvement of MCE can be achieved. Finally, it is worth noting that representing real molecular magnets by Ising models might not be realistic enough and considering corresponding Heisenberg models would be more appropriate. However, in spite of the expected differences, it has been shown that the step-wise structure of the magnetization in such small antiferromagnetic spin clusters as a function of the applied field observed in the Heisenberg models is qualitatively similar to that found in the corresponding Ising models~\cite{viit97a,viit97b,park00}. Nevertheless, the Heisenberg spin clusters have more degrees of freedom and for a fixed number of spins there are more distinct energy levels. Consequently, there are more steps in the zero-temperature magnetization process and the step widths related to the level-crossing transitions are typically smaller for the Heisenberg systems. From the MCE point of view an important consequence is the appearance of the zero-magnetization plateau in some Heisenberg systems, which in the Ising limit vanishes~\cite{karl17a,karl17b}. Therefore, it would be interesting to extend the present investigation to the corresponding Heisenberg systems and study the effects of the spin value and the single-ion anisotropy. These effects can partially suppress quantum fluctuations~\cite{chen13} and, consequently, also the appearance of the zero-magnetization plateau. Thus, in the proper setting it might be possible to restore the enhanced MCE even in such Heisenberg systems, at least in the highly anisotropic (semi-classical Ising) limit.

%
%
%

%

\vspace{6pt} 



\authorcontributions{Conceptualization, M.\v{Z}.; methodology, M.\v{Z}. and M.M.; software, M.M.; validation, M.\v{Z}. and M.M.; writing--original draft preparation, M.\v{Z}. and M.M.; writing--review and editing, M.\v{Z}. and M.M.; visualization, M.M.; supervision, M.\v{Z}. All authors have read and agreed to the published version of the manuscript.}

\funding{This research was funded by Vedeck\'{a} Grantov\'{a} Agent\'{u}ra M\v{S}VVa\v{S} SR a SAV (1/0531/19) | Agent\'{u}ra na Podporu V\'{y}skumu a V\'{y}voja (APVV-16-0186).}


\conflictsofinterest{The authors declare no conflict of interest.} 

\abbreviations{The following abbreviations are used in this manuscript:\\

\noindent 
\begin{tabular}{@{}ll}
MCE & magnetocaloric effect\\
T & triangular cluster\\
2CS & two corner-sharing tetrahedra cluster\\
GS & ground state
\end{tabular}}

\externalbibliography{yes}
\bibliography{bib_mce}



\end{document}